\documentclass[twocolumn]{aastex62}

\newcommand{\myemail}{speacock@lpl.arizona.edu}

\usepackage[T1]{fontenc}
\usepackage{romanbar}
\usepackage{appendix}
\usepackage{graphicx}

\accepted{April 13, 2020}

\submitjournal{ApJ}

\shorttitle{HAZMAT VI}
\shortauthors{Peacock et al.}

\begin{document}
\title{HAZMAT VI: The Evolution of Extreme Ultraviolet Radiation Emitted from Early M Stars}

\email{\myemail}

\author{Sarah Peacock}
\affil{University of Arizona, Lunar and Planetary Laboratory, 1629 E University Boulevard, Tucson, AZ 85721, USA}
\author{Travis Barman}
\affiliation{University of Arizona, Lunar and Planetary Laboratory, 1629 E University Boulevard, Tucson, AZ 85721, USA}
\author{Evgenya L. Shkolnik}
\affil{School of Earth and Space Exploration, Arizona State University, Tempe, AZ 85281, USA}
\author{R. O. Parke Loyd}
\affil{School of Earth and Space Exploration, Arizona State University, Tempe, AZ 85281, USA}
\author{Adam C. Schneider}
\affil{School of Earth and Space Exploration, Arizona State University, Tempe, AZ 85281, USA}
\author{Isabella Pagano}
\affil{INAF - Osservatorio Astrofisico di Catania, Via S. Sofia 78, 95123, Catania, Italy}
\author{Victoria S. Meadows}
\affil{Department of Astronomy and Astrobiology Program, University of Washington, Box 351580, Seattle, WA 98195, USA}
\affil{NASA NExSS Virtual Planetary Laboratory, Box 351580, University of Washington, Seattle, WA 98195, USA}

\begin{abstract}

Quantifying the evolution of stellar extreme ultraviolet (EUV, 100 -- 1000 \AA) emission is critical for assessing the evolution of planetary atmospheres and the habitability of M dwarf systems. Previous studies from the HAbitable Zones and M dwarf Activity across Time (HAZMAT) program showed the far- and near-UV (FUV, NUV) emission from M stars at various stages of a stellar lifetime through photometric measurements from the \textit{Galaxy Evolution Explorer} (\textit{GALEX}). The results revealed increased levels of short-wavelength emission that remain elevated for hundreds of millions of years. The trend for EUV flux as a function of age could not be determined empirically because absorption by the interstellar medium prevents access to the EUV wavelengths for the vast majority of stars. In this paper, we model the evolution of EUV flux from early M stars to address this observational gap. We present synthetic spectra spanning EUV to infrared wavelengths of 0.4 $\pm$ 0.05 $M_\sun$ stars at five distinct ages between 10 and 5000 Myr, computed with the PHOENIX atmosphere code and guided by the \textit{GALEX} photometry. We model a range of EUV fluxes spanning two orders of magnitude, consistent with the observed spread in X-ray, FUV, and NUV flux at each epoch. Our results show that the stellar EUV emission from young M stars is 100 times stronger than field age M stars, and decreases as $t^{-1}$ after remaining constant for a few hundred million years. This decline stems from changes in the chromospheric temperature structure, which steadily shifts outward with time. Our models reconstruct the full spectrally and temporally resolved history of an M star's UV radiation, including the unobservable EUV radiation, which drives planetary atmospheric escape, directly impacting a planet's potential for habitability.

\end{abstract}


\NewPageAfterKeywords

\section{Introduction}\label{sec:intro}
M stars are the most common spectral type in the galaxy \citep{Reid1997} and offer many observational advantages for detecting and characterizing potentially habitable planets; however, their extended pre-main-sequence phase \citep{Tian2015} and accompanying high levels of short wavelength radiation \citep{Shkolnik2014,Schneider2018} can alter or destroy a planet's atmosphere and leave it uninhabitable. The close proximity of planets located within the canonical habitable zone of M stars (0.1 -- 0.4 au) corresponds to increased exposure to high-energy radiation, which is further elevated during the first several hundred million years while planets are forming their primary and, in some cases, secondary atmospheres \citep{Shkolnik2014}. Since short-wavelength (X-ray, 5 -- 100 \AA; EUV, 100 -- 1000 \AA; FUV, 1100 -- 1800 \AA; NUV, 1800 -- 3100 \AA) radiation drives erosion and chemical modification of planetary atmospheres \citep{Segura2005,Lammer2007,Tian2008,Koskinen2010,Segura2010,Moses2014,Chad2015,Rugheimer2015,Luger2015}, it is crucial to assess the complete history of the UV environment around M stars in order to better understand the early evolution of the protoplanetary disk, the stability of close-in planet atmospheres, and the location and evolution of the habitable zone.

As FGKM stars spin down with age, the dynamo production of magnetic fields reduces and the chromospheric activity of these stars decreases. The Sun in Time program \citep{Dorren1994,Guinan2004} determined the spectral irradiance of the Sun over its main-sequence lifetime from multi-wavelength observations of G-type solar proxies at different ages and found that the combined X-ray and EUV (XUV, 5 -- 1000 \AA) emission from the young Sun was 100 -- 1000 times stronger than that of the present-day \citep{Ribas2005}. The evolution of X-ray, FUV, and NUV emission from early -- mid M stars has also already been observed and characterized as a part of the HAbitable Zones and M dwarf Activity Across Time (HAZMAT) program \citep{Shkolnik2014}. Similar to the evolutionary trends observed in G-type stars, \cite{Shkolnik2014} found that M stars emit higher levels of short wavelength radiation at young ages, showing that the median fractional \textit{R\"ontgensatellit} (\textit{ROSAT}) X-ray, and \textit{GALEX} FUV and NUV fluxes all remain at a constant level for a few hundred million years followed by a decline in activity that goes as $\sim t^{-1}$. The median fractional flux reduces by factors of 30 and 20 for FUV and NUV respectively, as compared to a factor of 65 for X-ray. The evolutionary trend for EUV flux for M stars has not yet been quantified.

Assessing the lifetime exposure of habitable zone planets to stellar EUV radiation is critical for understanding the evolution of planetary atmospheres, as thermal and non-thermal atmospheric escape rates are highest during the early active phase of the host star, before 500 Myr \citep{Lammer2012}. Studies of terrestrial atmospheres exposed to the increased amount of XUV emission expected from the young Sun suggest that the development of life on Earth or Mars could have been strongly influenced by the evolutionary history of the Sun's high-energy radiation \citep{Cockell2000}. Similar to the strong UV environment of the early solar system, elevated levels of XUV emission from young M stars could affect the evolution of atmospheres and climates on potentially habitable exoplanets.

Unfortunately, there are no currently operational telescopes that can observe M stars across EUV wavelengths. There were few M stars observed with the \textit{Far Ultraviolet Spectroscopic Explorer} (\textit{FUSE}; 920 -- 1180 \AA) and archival EUV spectra from the \textit{Extreme Ultraviolet Explorer} (\textit{EUVE}; 100 -- 400 \AA) exist for six actively flaring M stars all within 10 pc. This small sample does not provide a comprehensive view of the evolution of EUV flux emitted from M stars. Furthermore, contamination from optically thick interstellar hydrogen obscures the 400 -- 912 \AA \ region of the EUV spectrum. While this portion of the spectrum will always be inaccessible due to this interstellar contamination, a future large-area EUV dedicated mission would yield better quality spectra than the \textit{EUVE} spectrometers (effective area $\sim$1 cm$^2$) and could provide much needed observations for the quiet M stars that are potentially (or likely) less hazardous for habitable worlds than more active M stars. A dedicated 1 m telescope with an assumed 30\% throughput could measure the EUV spectrum for an average M star within 10 pc at a signal-to-noise ratio of 100 (2 hour integration time). Within 10 pc, there are nearly 300 M dwarfs, 25 of which host confirmed exoplanets\footnote{Data obtained from the NASA Exoplanet Archive, https://exoplanetarchive.ipac.caltech.edu/}. 

Due to these observational restrictions, the scientific community must rely on either empirical scaling relationships or a limited number of synthetic spectra to quantify EUV flux. Since the radiation in individual EUV emission lines penetrates planetary atmospheres at different depths, affecting the ionization rates and likelihood for escape, it is necessary to include high-resolution EUV spectra rather than single-valued fluxes when modeling both the photochemistry and escape in exoplanet atmospheres. To calculate stellar EUV spectra, semiempirical models are guided by and validated using X-ray and/or UV observations. The fluxes in these wavelength regimes emerge from different depths in a stellar atmosphere providing important information about the temperature structure. Both FUV and NUV continua and emission lines form in the chromosphere and transition region (TR) at temperatures between 10$^4$ and 10$^6$ K \citep{Sim2005}, while X-ray photons originate in the corona ($>$10$^6$ K). Semiempirical modeling efforts have shown that the EUV spectrum is generated from photons emerging from the full upper atmosphere \citep{Fontenla2016}, with the majority of the EUV continuum and emission lines forming at temperatures below 2$\times$10$^5$ K \citep{Peacock2019b}. 

From \cite{Shkolnik2014}, there are both X-ray and FUV--NUV observations available for M stars spanning a wide range of ages that can be used to guide semiempirical models and compute EUV spectra of M stars at various epochs. Models based on non-contemporaneous X-ray and UV observations, however, will potentially have large uncertainties in the predicted EUV spectra as a result of the highly variable levels of measured X-ray and UV flux caused by flares and magnetic activity occurring in the upper atmospheric layers of M stars \citep{Monsignori1996,Haw2003,Stelzer2013,Loyd18a,Loyd2018b}. Since X-ray observations do not probe beneath the stellar corona, models fit solely to X-ray detections underestimate UV (100 -- 3000 \AA) fluxes in known planet hosts due to the lack of contribution from the deeper stellar layers \citep{Linsky2014,Louden2017}.

In this paper, we model the evolution of EUV flux from early M stars from 10 Myr to 5 Gyr. We construct upper atmosphere models using empirical guidance from \textit{GALEX} FUV and NUV photometry to produce high-resolution ($\lambda <$ 0.1 \AA) 1D nonlocal thermodynamic equilibrium (non-LTE) synthetic spectra (EUV--IR, 100 \AA\ -- 5.5 $\mu$m) and quantify how the temperature structure for an average early M star evolves with time. We present sets of synthetic spectra of 0.4 $\pm$ 0.05 $M_\sun$ stars that reproduce the range of \textit{GALEX} measurements of 10, 45, 120, 650 Myr, and 5 Gyr early M stars. We also validate the synthetic spectra representative of 45, 650 Myr, and 5 Gyr stars using \textit{Hubble Space Telescope} (\textit{HST}) UV spectra of 45 Myr Tuc-Hor members, 650 Myr Hyades members, and field stars.

\section{Model Construction and Parameterization}\label{sec:model}

The PHOENIX atmosphere code \citep{Hauschildt1993, Hauschildt2006, Baron2007} has the flexibility to model atmospheres and compute synthetic spectra for stars of all masses and temperatures, and is a leading code for the modeling of main-sequence stars \citep{Hauschildt1999}, brown dwarfs \citep{Allard2001}, white dwarfs \citep{Barman2000}, and giants \citep{Aufdenberg2002}. Recent chromospheric modeling studies have used PHOENIX to compute synthetic spectra of M-type stars from 100 \AA \ to 5 $\mu$m and have successfully reproduced observations of planet-hosting M stars (M0 -- M8; 0.5 -- 0.08 $M_\sun$) at near-IR, optical, NUV, and FUV wavelengths \citep{Hintz2019,Peacock2019b,Peacock2019}.

\begin{table}[t!]
    \centering
    \begin{tabular}{lccc|ccc}
    \hline\hline
    \textbf{$M_\star$}&\multicolumn{3}{c}{\textbf{0.45 $M_\sun$}} \vline &\multicolumn{3}{c}{\textbf{0.35 $M_\sun$}}\\
          \hline 
        Age & $T_{\rm eff}$ & log($g$) & $R_\star$ & $T_{\rm eff}$ & log($g$) & $R_\star$\\
        (Myr)  & (K) & (cm s$^{-2}$) & ($R_\sun$) & (K) & (cm s$^{-2}$) & ($R_\sun$)   \\
        \hline
        10\dotfill &  3550 & 4.24 & 0.84 & 3400 & 4.22 & 0.76\\
        45\dotfill & 3550 & 4.6   & 0.55 & 3400 & 4.6 & 0.48\\
        120\dotfill & 3600 & 4.8  & 0.44 & 3450 & 4.85 & 0.36\\
        650\dotfill & 3600 & 4.85  & 0.42 & 3450 & 4.95 & 0.33\\
        5000 & 3600 & 4.85 & 0.42 & 3450 & 4.95 & 0.33\\
        \hline
    \end{tabular}
    \caption{Stellar parameters used for the models. The BHAC15 models \citep{Baraffe2015} were used to obtain the values, taking a rounded average between 0.5 $M_\sun$ and 0.4 $M_\sun$ stars for the 0.45 $M_\sun$ models, and 0.3 and 0.4 $M_\sun$ stars for the 0.35 $M_\sun$ models.}
    \label{tab:ageparams}
\end{table}

We use PHOENIX to construct grids of 1D upper atmosphere models of 0.45 $M_\sun$ and 0.35 $M_\sun$ stars at five distinct ages between 10 and 5000 Myr. These stellar masses are characteristic of M1 -- M2 stars with solar-like metallicities of [M/H] $\approx$ 0 to -0.5 \citep{Baraffe1996} and represent stars that are partially convective ($\geq$ 0.35 $M_\sun$). The models are computed in hydrostatic equilibrium on a log(column mass) grid. We begin with a photosphere model in radiative-convective equilibrium computed with the effective temperature ($T_{\rm eff}$) and surface gravity ($g$) corresponding to the stellar mass at a particular age. We use the \cite{Baraffe2015} BHAC15 models to determine these values at 10, 45, 120, 650 Myr, and 5 Gyr (Table \ref{tab:ageparams}). 

\begin{figure}[t!]
    \centering
    \includegraphics[width=1.0\linewidth]{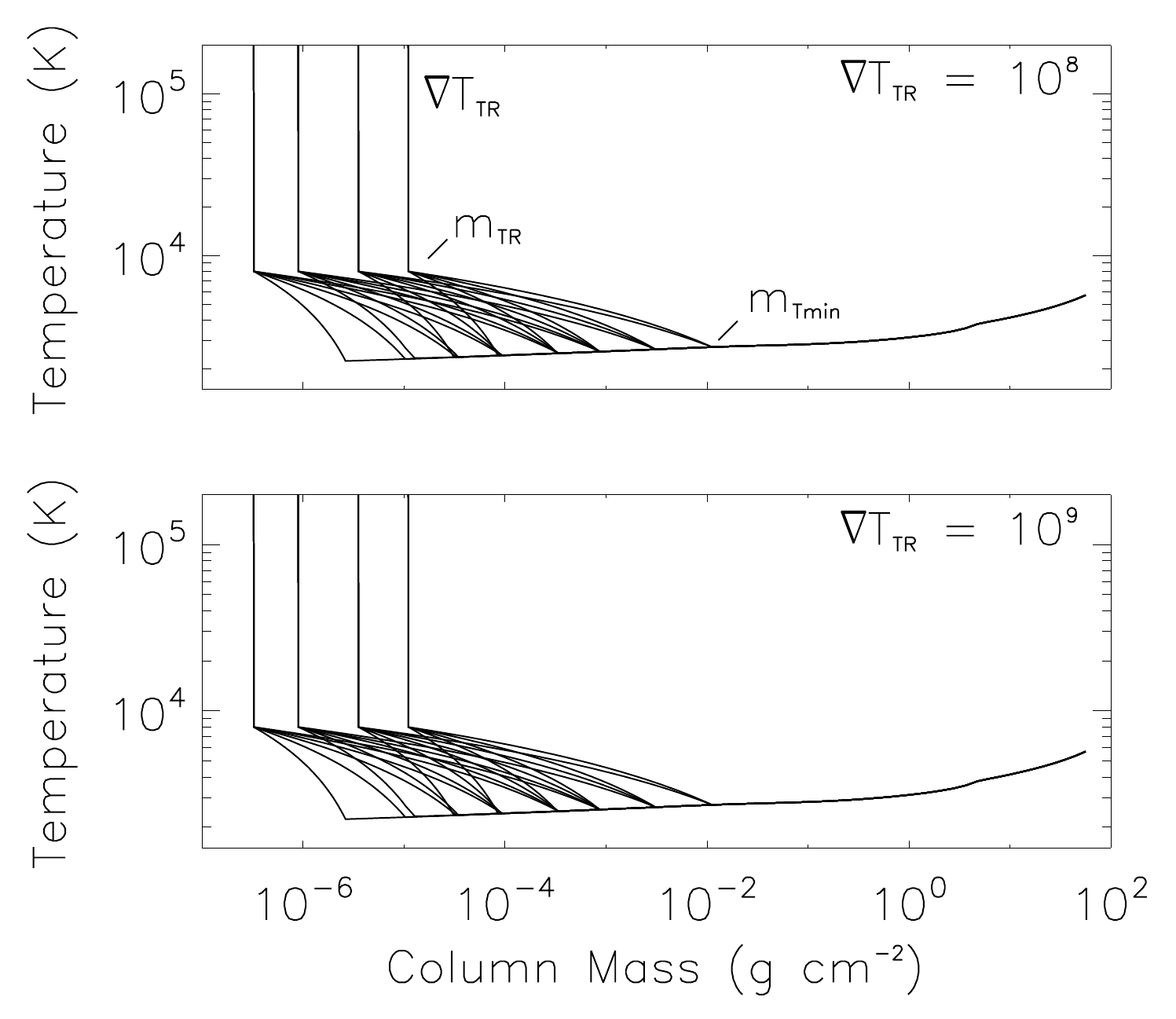}
    \caption{Temperature structures of models for a $\geq$200 Myr 0.45 $M_\sun$ star. Upper panel: series of varying log \textit{m}$_{\rm TR}$ and log \textit{m}$_{\rm Tmin}$ with $\nabla T_{\rm TR}$ = 10$^8$ K dyne$^{-1}$ cm$^{2}$. Lower panel: series of varying log \textit{m}$_{\rm TR}$ and log \textit{m}$_{\rm Tmin}$ with $\nabla T_{\rm TR}$ = 10$^9$ K dyne$^{-1}$ cm$^{2}$. In both series, log \textit{m}$_{\rm TR}$ varies from -6.5 g cm$^2$ -- -5 g cm$^{-2}$ and log \textit{m}$_{\rm Tmin}$ varies from -5.5 to -2 g cm$^{-2}$.}
    \label{fig:cmtall}
\end{figure}

Building upon the underlying photosphere models, we compute a series of 46 systematically varying upper atmospheric temperature structures using a similar prescription to that used in \cite{Peacock2019b,Peacock2019}. The upper atmospheres consist of temperature distributions for a chromosphere increasing up to 8000 K, followed by a steep temperature gradient in the TR extending to a maximum temperature of 2 $\times$ 10$^5$ K. The models explore a wide range of chromospheric pressures and thicknesses, and two values of TR thickness simulating a range of UV activity levels observed in early M stars of all ages. We implement linear rises in temperature-log(column mass) in both regions and modify three free parameters: the column mass at the initial chromospheric temperature rise (\textit{m$_{\rm Tmin}$}), the column mass at the top of the chromosphere (\textit{m$_{\rm TR}$}), and the temperature gradient in the TR ($\nabla T_{\rm TR}$ = $|$d \textit{T}/d log \textit{P}$|$). In Figure \ref{fig:cmtall}, we show the complete grid of models for a 0.45 $M_\sun$ star greater than 200 Myr old. For each stellar mass and age\footnote{In the \cite{Baraffe2015} BHAC15 models, $T_{\rm eff}$ and log(\textit{g}) remain nearly constant for a given mass at ages beyond 200 Myr, so a singular grid of models is used to identify realistic spectra of 650 Myr and 5 Gyr.}, we explore models with log \textit{m}$_{\rm Tmin}$~=~-5.5 -- -2~g~cm$^{-2}$ and log \textit{m}$_{\rm TR}$ = -6.5 -- -5 g~cm$^{-2}$ in step sizes of 0.5 g~cm$^{-2}$, and $\nabla T_{\rm TR}$~=~10$^8$, 10$^9$~K~dyne$^{-1}$~cm$^{2}$. 

The physical conditions in stellar upper atmospheres are such that radiative rates are much larger than collisional rates and radiative transfer is dominated by non-LTE effects. For our models, we do multi-line non-LTE calculations for the 23 most common elements found in the Sun. We consider a total of 15,355 levels and 233,871 emission lines when computing our set of 73 atoms and ions: \ion{H}{1}, He {\small\rmfamily I -- II\relax}, C {\small\rmfamily I -- IV\relax} N {\small\rmfamily I -- IV\relax}, O {\small\rmfamily I -- IV\relax}, Ne {\small\rmfamily I -- II\relax}, Na {\small\rmfamily I -- III\relax}, Mg {\small\rmfamily I -- IV\relax}, Al {\small\rmfamily I -- III\relax}, Si {\small\rmfamily I -- IV\relax}, P {\small\rmfamily I -- II\relax}, S {\small\rmfamily I -- III\relax}, Cl {\small\rmfamily I -- III\relax}, Ar {\small\rmfamily I -- III\relax}, K {\small\rmfamily I -- III\relax}, Ca {\small\rmfamily I -- III\relax}, Ti {\small\rmfamily I -- IV\relax}, V {\small\rmfamily I -- III\relax}, Cr {\small\rmfamily I -- III\relax}, Mn {\small\rmfamily I -- III\relax}, Fe {\small\rmfamily I -- VI\relax}, Co {\small\rmfamily I -- III\relax}, and Ni {\small\rmfamily I -- III\relax}.

We use the complete frequency redistribution (CRD) approximation to calculate the majority of our line profiles, as CRD accounts for overlapping radiative transitions and is generally a good approximation. In some strong resonance lines, however, the CRD approximation has been shown to inaccurately account for coherent scattering of photons and can result in calculated line profiles with overpredicted wings \citep{Hubeny1995,Uitenbroek2001,Peacock2019}. PHOENIX is equipped with the partial frequency redistribution (PRD) capabilities necessary to accurately compute the radiative losses in these lines and is included when computing both \ion{H}{1} Lyman $\alpha$ (1215.7 \AA) and \ion{Mg}{2} \textit{h} (2802.7 \AA) and \textit{k} (2795.53 \AA). These emission features are the two strongest in the UV spectrum and are commonly used chromospheric diagnostics.

\begin{figure*}[t!]
    \centering
    \includegraphics[scale=1.1]{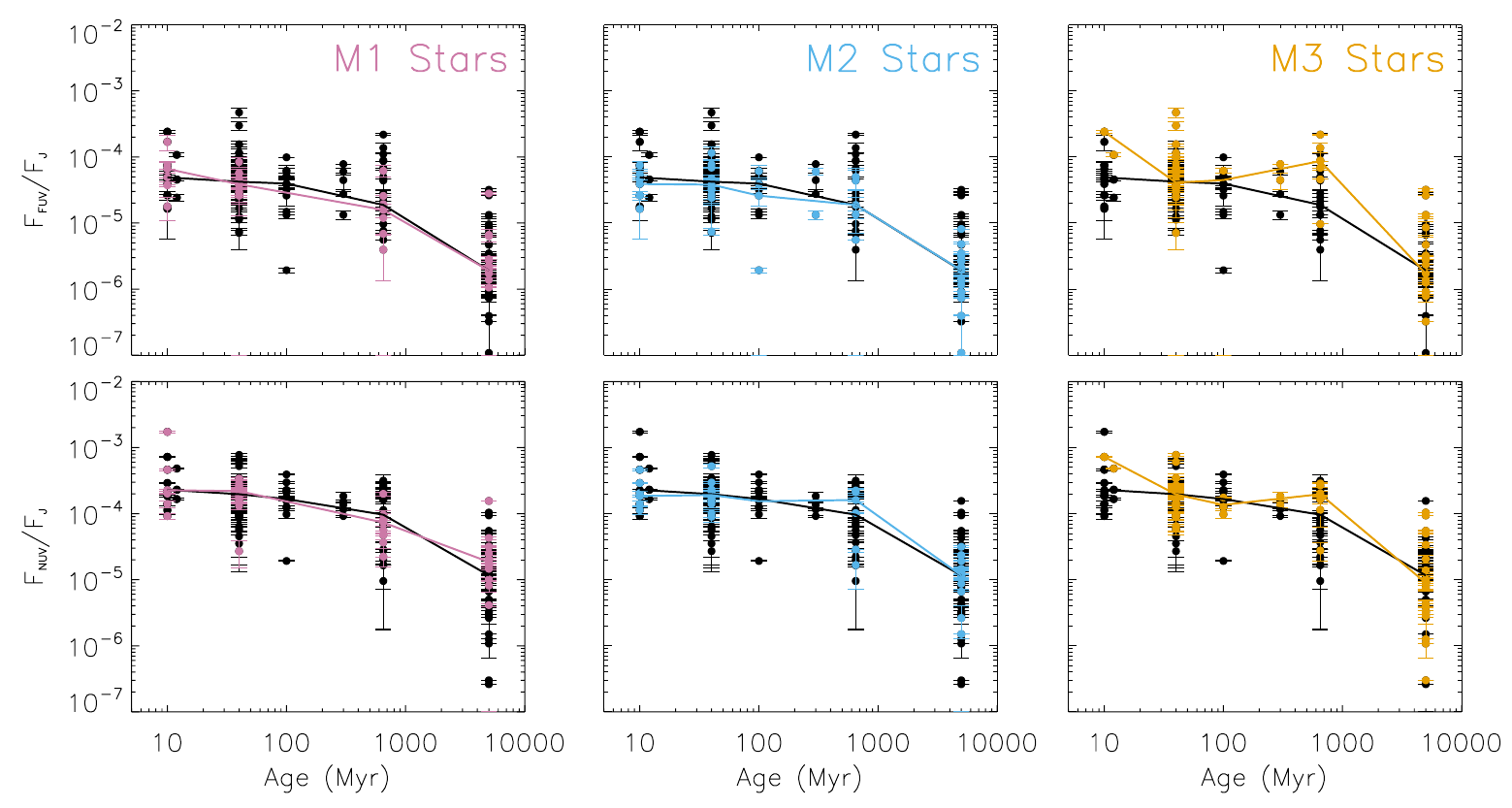}
    \caption{Fractional UV flux densities for the early -- mid M stars in the HAZMAT I sample as a function of stellar age \citep{Shkolnik2014}. FUV upper limits are estimated from the NUV flux density using a power-law coefficient of 1.11. The full sample of K7 -- M4 stars are plotted in black, M1 stars are plotted in purple, M2 stars in blue, and M3 stars in yellow. The median fluxes per sample are connected by lines in the corresponding color. The median fractional UV flux densities of M1 and M2 stars closely match the median values for the full sample, while the median values for the M3 stars deviate at both 10 Myr and 650 Myr due to small sample size and a large fraction of nondetections in the FUV channel.}
    \label{fig:ExFlux}
\end{figure*}

\section{The HAZMAT Stellar Sample}\label{sec:sample}

As part of the HAZMAT program, \cite{Shkolnik2014} (HAZMAT I) and \cite{Schneider2018} (HAZMAT III) assessed the comprehensive FUV and NUV history of a statistical sample of K7--M8 stars ranging in age from tens to hundreds to thousands of megayears. The early -- mid M stars ($<$M4) in the HAZMAT~I sample include members of the 10 Myr TW Hydra, 24 Myr $\beta$ Pic, 45 Myr Tuc-Hor, 120 Myr AB Dor, 300 Myr Ursa Major, and 650 Myr Hyades young moving groups (YMGs) as well as an old population of M stars with an average age of $\sim$5 Gyr. 

For this study, we present models representative of M1--M2-type stars (0.4 $\pm$ 0.05 $M_\sun$\footnote{Spectral type-mass estimates are based on Table 1 of \cite{Schneider2018}.}) at 10 Myr, 45 Myr, 120 Myr, 650 Myr, and 5 Gyr. From our series of upper atmosphere models, we identify a set of five synthetic spectra per age and stellar mass (0.45 $M_\sun$, 0.35 $M_\sun$) that represent the range of observed UV flux, as guided by the FUV and NUV \textit{GALEX} photometry of the stars in the HAZMAT I paper. We group data from the 24 Myr $\beta$ Pic members with those of the 10 Myr TW Hydra association, and group data from the 300 Myr Ursa Major stars with the 120 Myr AB Dor sample. Where available, we validate that the synthetic spectra match \textit{HST} UV observations of stars of corresponding age. All target stars were identified as members of each group using 3D space velocities and other age indicators including H$\alpha$ emission, lithium absorption, and low gravity indices \citep{Shkolnik2009, Shkolnik2011,Shkolnik2012, Kraus2014}.

\newpage

\subsection{\textit{GALEX} FUV and NUV Photometry}

\textit{GALEX} FUV (1340--1810 \AA, $\lambda_{\rm eff}$=1542 \AA) and NUV (1687--3010 \AA, $\lambda_{\rm eff}$=2274 \AA) flux densities measured for the 215 K7--M4 stars (0.5 -- 0.2 $M_\sun$) in our reference sample are listed in Table 1 of \cite{Shkolnik2014} along with their published spectral types and Two Micron All Sky Survey (2MASS) \textit{J} magnitudes. Since trigonometric parallaxes were not available for all of the stars at the time of the \cite{Shkolnik2014} study, fractional UV flux densities relative to \textit{J}$_{\rm 2MASS}$ flux densities were used. We adopt those measurements here.

The FUV and NUV fractional flux densities plotted as a function of age for the full stellar sample are presented in black in Figure \ref{fig:ExFlux}. Ninety-five percent of the targets observed have \textit{GALEX} NUV detections, while 68\% of the 184 stars were detected in the FUV channel (31 stars were not observed in the FUV). In the cases where stars have published FUV upper limits in place of detections, we estimate the excess FUV flux density using a power-law relationship,

\begin{equation}
  (F_{\rm FUV}/F_{\rm J})_{exc} \propto (F_{\rm NUV}/F_{\rm J})_{exc}^{1.11},  
\end{equation}

\noindent derived in \cite{Shkolnik2014} from a regression fit to the FUV and NUV detections of the stellar sample. In Figure \ref{fig:ExFlux}, the median excess flux density per age is connected with a black line and shows a clear decrease in age for both FUV and NUV emission, with a steep drop-off after 650 Myr. Within each age group, there is a 1 -- 2 order of magnitude spread in excess UV flux likely due to flares, uncertainty in ages, and a large spread in measured rotation periods.

\begin{figure*}[t!]
    \centering
    \includegraphics[width=1.0\textwidth]{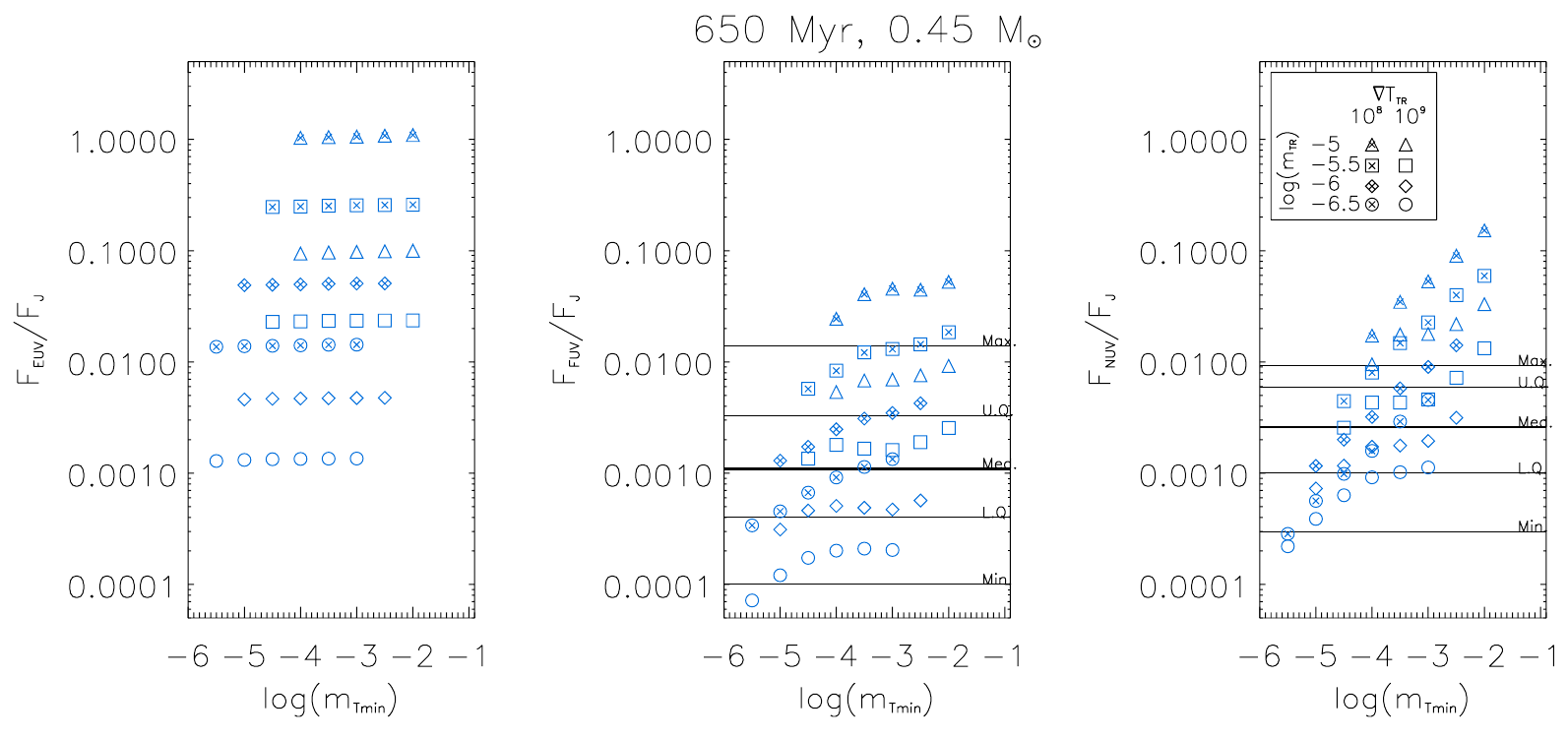}
    \caption{EUV (left), FUV (middle), and NUV (right) fractional flux densities for the suite of upper atmosphere models representative of $\geq$200 Myr 0.45 $M_\sun$ stars as a function of the model parameter, $m_{\rm Tmin}$. The shapes represent models with a particular $m_{\rm TR}$ (see the legend in the right panel). Empty shapes correspond to models with $\nabla T_{\rm TR}$ = $10^8$ K dyne$^{-1}$ cm$^2$, filled shapes correspond to models with $\nabla T_{\rm TR}$ = $10^9$ K dyne$^{-1}$ cm$^2$. The statistical quantities (minimum, lower quartile, median, upper quartile, and maximum) for the 650 Myr HAZMAT sample are indicated with black horizontal lines. The PHOENIX models identified as the matches to each FUV--NUV quantity pair are listed in Table \ref{tab:fluxes}.}
    \label{fig:minmax}
\end{figure*}

The most common subtypes represented at each age in the HAZMAT I sample are M1, M2, and M3. In Figure \ref{fig:ExFlux}, we analyze the evolutionary trends for these subtypes as compared to the total sample and find that the median excess UV fluxes for the M1 and M2 stars closely replicate those for the full sample, while the M3 stars deviate at 10 and 650 Myr. The minimum, median, and maximum excess UV flux densities per subtype are listed in Table \ref{tab:app_ageparams} of Appendix \ref{app:minmax} along with the sample size (separated into number of detections and FUV upper limits) per age group. The increased median excess flux densities for the 10 Myr M3 stars are likely due to the small sample size, of which two of the three stars are visual binaries. A large percentage (53\%) of FUV upper limits in the 650 Myr Hyades sample may be influencing the median value determined for this age group. Since both the median and spread of each subtype are generally consistent when not influenced by sample size, we compare both the 0.45 $M_\sun$ and 0.35 $M_\sun$ models to the fluxes of the full sample of K7--M4 stars rather than separating by subtype.

\subsection{\textit{HST} FUV and NUV Spectra}

In addition to the large sample of \textit{GALEX} broadband FUV and NUV photometry, the HAZMAT program obtained \textit{HST} Cosmic Origins Spectrograph (COS) spectra (PID \#14784, PI Shkolnik) for a sample of 12 M0 -- M2 members in the 45 Myr Tuc-Hor association, five M1 -- M2 members in the 650 Myr Hyades cluster, and three M0 -- M2 field age stars. 

The observations were taken using the G230L, G230M, and G130M gratings on COS and covered wavelength regions between 1170 and 3200 \AA. The Tuc-Hor targets were observed over two orbits, the Hyades targets were observed over eight orbits, and the field stars over three orbits. For each target, the NUV continuum between $\sim$2800 and 3200 \AA \ and several FUV and NUV emission lines were detected, including \ion{C}{3} (1176 \AA), \ion{Si}{3} (1206 \AA), \ion{H}{1} Lyman $\alpha$ (1216 \AA), \ion{N}{5} (1239, 1243 \AA), \ion{C}{2} (1337 \AA), \ion{Si}{4} (1393 \AA, 1402 \AA), \ion{C}{4} (1548, 1550 \AA), \ion{O}{2} (1639 \AA), and \ion{Mg}{2} \textit{h} and \textit{k} (2796 \AA, 2803 \AA). 

Supplementing this sample, \textit{HST} COS and Space Telescope Imaging Spectrograph (STIS) UV spectra for 11 old low-mass stars (seven M and four K dwarfs) are also available, as collected as a part of the MUSCLES Treasury Survey \citep{France2016}. The stars in the MUSCLES sample are exoplanet hosts and likely biased toward lower activity levels since planet surveys select for quiescent stars. In a recent paper, we calculated synthetic EUV--IR spectra that closely reproduce the FUV and NUV \textit{HST} spectra of three of the stars from this survey  \citep{Peacock2019b}.

\section{Analysis}\label{sec:analysis}

Within each age group in the HAZMAT I sample, there is a 1 -- 2 order of magnitude range in both the FUV and NUV excess fractional flux densities. To capture the intrinsic variability of short-wavelength emission from M stars, we present sets of models representative of 0.4 $\pm$ 0.05 $M_\sun$ stars that reproduce the full spread of \textit{GALEX} measurements at each age. 

\begin{table*}[th!]
    \centering
    \begin{tabular}{lc|cc|cc}
    \hline \hline
   & \textbf{$F_{\rm EUV}$/$F_{\rm J} \times 10^3$}& \multicolumn{2}{c}{\textbf{$F_{\rm FUV}$/$F_{\rm J} \times 10^3$}} & \multicolumn{2}{c}{\textbf{$F_{\rm NUV}$/$F_{\rm J} \times 10^3$}}\\
    \hline
    & Model &  Model & \textit{GALEX} &  Model & \textit{GALEX}\\
    \hline
    \textbf{10 Myr} \\ \hline
     Maximum\dotfill        & 393.3	$\pm$	106.7	&21.3	$\pm$	5.0	&	15.4	$\pm$	1.0	&38.9	$\pm$	2.2	&	51.1	$\pm$	2.1	\\
     Upper Quartile\dotfill & 56.3	$\pm$	11.2	&4.4	$\pm$	1.0	&	4.8	$\pm$	0.7	&14.4	$\pm$	0.1	&	13.6	$\pm$	0.5	\\
     Median\dotfill         & 35.5	$\pm$	10.5	&2.8	$\pm$	0.6	&	2.9	$\pm$	0.1	&6.2	$\pm$	0.9	&	6.7	$\pm$	0.1	\\
     Lower Quartile\dotfill & 22.8	$\pm$	8.2	&1.8	$\pm$	0.4	&	1.7	$\pm$	0.3	&4.5	$\pm$	0.2	&	5.2	$\pm$	0.2	\\
     Minimum\dotfill        & 11.6	$\pm$	2.9	&1.0	$\pm$	0.2	&	1.1	$\pm$	0.4	&2.4	$\pm$	0.2	&	2.7	$\pm$	0.3	\\
    \hline
    \textbf{45 Myr }\\ \hline
     Maximum\dotfill        & 658.1	$\pm$	224.3	&23.0	$\pm$	0.3	&	30.2	$\pm$	5.0	&19.7	$\pm$	3.2	&	22.8	$\pm$	1.0	\\
     Upper Quartile\dotfill & 46.1	$\pm$	4.3	&3.3	$\pm$	0.3	&	3.6	$\pm$	0.5	&7.8	$\pm$	0.9	&	7.3	$\pm$	0.2	\\
     Median\dotfill         & 32.0	$\pm$	9.8	&2.5	$\pm$	0.5	&	2.6	$\pm$	0.4	&6.0	$\pm$	1.0	&	5.8	$\pm$	0.2	\\
     Lower Quartile\dotfill & 26.1	$\pm$	3.9	&2.1	$\pm$	0.1	&	1.9	$\pm$	0.6	&4.9	$\pm$	0.1	&	4.2	$\pm$	0.3	\\
     Minimum\dotfill        & 6.6	$\pm$	2.0	&0.5	$\pm$	0.1	&	0.4	$\pm$	0.1	&1.0	$\pm$	0.1	&	0.8	$\pm$	0.4	\\
    \hline
    \textbf{120 Myr} \\ \hline
     Maximum\dotfill        & 49.3	$\pm$	2.8	&3.63	$\pm$	0.03	&	4.0	$\pm$	0.9	    &11.4	$\pm$	1.9	&	11.6	$\pm$	0.3	\\
     Upper Quartile\dotfill & 50.5	$\pm$	1.5	&3.2	$\pm$	0.1	&	3.0	$\pm$	0.4	        &6.0	$\pm$	0.2	&	6.1	$\pm$	0.3	\\
     Median\dotfill         & 22.9	$\pm$	0.5	&1.80	$\pm$	0.01	&	2.0	$\pm$	0.3	    &4.6	$\pm$	0.1	&	4.7	$\pm$	0.1	\\
     Lower Quartile\dotfill & 12.5	$\pm$	9.9	&1.6	$\pm$	0.1	&	1.6	$\pm$	0.1	        &4.23	$\pm$	0.03	&	3.7	$\pm$	0.1	\\
     Minimum\dotfill        & 1.4	$\pm$	0.1	&0.12	$\pm$	0.01	&	0.1	$\pm$	0.01	&0.39	$\pm$	0.02	&	0.6	$\pm$	0.01	\\
    \hline
    \textbf{650 Myr} \\ \hline
     Maximum\dotfill        & 238.9	$\pm$	11.1	&9.6	$\pm$	1.3	&	14.0	$\pm$	0.4	&11.0	$\pm$	3.0	&	9.3	$\pm$	2.0	\\
     Upper Quartile\dotfill & 50.5	$\pm$	0.1	&3.1	$\pm$	0.1	&	3.3	$\pm$	1.2	        &5.74	$\pm$	0.04	&	5.9	$\pm$	0.6	\\
     Median\dotfill         & 21.9	$\pm$	1.0	&1.4	$\pm$	0.1	&	1.1	$\pm$	0.2	        &2.6	$\pm$	0.1	&	2.6	$\pm$	0.1	\\
     Lower Quartile\dotfill & 4.69	$\pm$	0.02&0.46	$\pm$	0.01	&	0.4	$\pm$	0.1	    &1.2	$\pm$	0.1	&	1.0	$\pm$	0.4	\\
     Minimum\dotfill        & 7.6	$\pm$	6.1	&0.2	$\pm$	0.1	&	0.1	$\pm$	0.1	&0.4	$\pm$	0.1	&	0.3	$\pm$	0.2	\\
    \hline
    \textbf{5 Gyr} \\ \hline
     Maximum\dotfill        & 22.2	$\pm$	0.9	&1.63	$\pm$	0.18	&	2.05	$\pm$	1.08	&4.37	$\pm$	0.05	&	4.59	$\pm$	0.05	\\
     Upper Quartile\dotfill & 3.1	$\pm$	1.5	&0.25	$\pm$	0.06	&	0.27	$\pm$	0.05	&0.69	$\pm$	0.04	&	0.78	$\pm$	0.004	\\
     Median\dotfill         & 1.4	$\pm$	0.1	&0.13	$\pm$	0.01	&	0.12	$\pm$	0.02	&0.40	$\pm$	0.01	&	0.42	$\pm$	0.3	\\
     Lower Quartile\dotfill & 1.4	$\pm$	0.1	&0.08	$\pm$	0.01	&	0.09	$\pm$	0.01	&0.23	$\pm$	0.01	&	0.27	$\pm$	0.08	\\
     Minimum\tablenotemark{a}\dotfill& 0.017	$\pm$	0.002	&0.011	$\pm$	0.003	&	0.01	$\pm$	0.02	&0.14	$\pm$	0.01	&	0.08	$\pm$	0.02	\\
    \hline
    \end{tabular}
    \caption{Fractional flux densities (multiplied by 10$^3$ for clarity) of synthetic spectra representative of 0.4 $\pm$ 0.05 $M_\sun$ stars at various ages. The models match paired FUV--NUV statistical quantities spanning the range of \textit{GALEX} measurements at each age.\\
    $^a$ Within the original explored parameter space, none of the $\geq$200 Myr models matched the minimum 5 Gyr excess FUV and NUV flux densities. We computed additional models with $\nabla T_{\rm TR}$ = 10$^{10}$ and 10$^{11}$ , log($m_{\rm TR}$) = -6.5, and log($m_{\rm Tmin}$) = -5.5, of which the $\nabla T_{\rm TR}$ = 10$^{11}$ model more closely reproduces the minimum \textit{GALEX} detections.}
    \label{tab:fluxes}
\end{table*}

For each one of our 386 computed spectra, we calculate synthetic photometry over the same wavelengths as the \textit{GALEX} FUV and NUV filter profiles and the \textit{J}$_{\rm 2MASS}$ band filter profile. In Figure \ref{fig:minmax}, we plot the EUV, FUV, and NUV fractional flux densities for the grid of upper atmosphere models representative of $\geq$200~Myr 0.45~$M_\sun$ stars. Across the entire UV spectrum, higher fluxes are generated by employing a more shallow temperature gradient in the TR and by shifting the top of the chromosphere inwards. A decrease in $\nabla T_{\rm TR}$ or increase in $m_{\rm TR}$ by factors of 10 results in nearly an order of magnitude increase in fractional EUV -- NUV flux density, with shorter wavelengths showing increasingly more sensitivity to these parameters. The EUV spectrum is very sensitive to $\nabla T_{\rm TR}$ and $m_{\rm TR}$, but is unaffected by changes in $m_{\rm Tmin}$. FUV flux densities are more sensitive to changes in $m_{\rm TR}$ than $m_{\rm Tmin}$, but do show a dependence on all three parameters. Similarly, the NUV spectrum is effected by changes in all three parameters, with changes in $m_{\rm Tmin}$ having the largest effect on these wavelengths as compared to the the EUV and FUV.

\begin{figure*}[t!]
    \centering
    \includegraphics[scale=0.85]{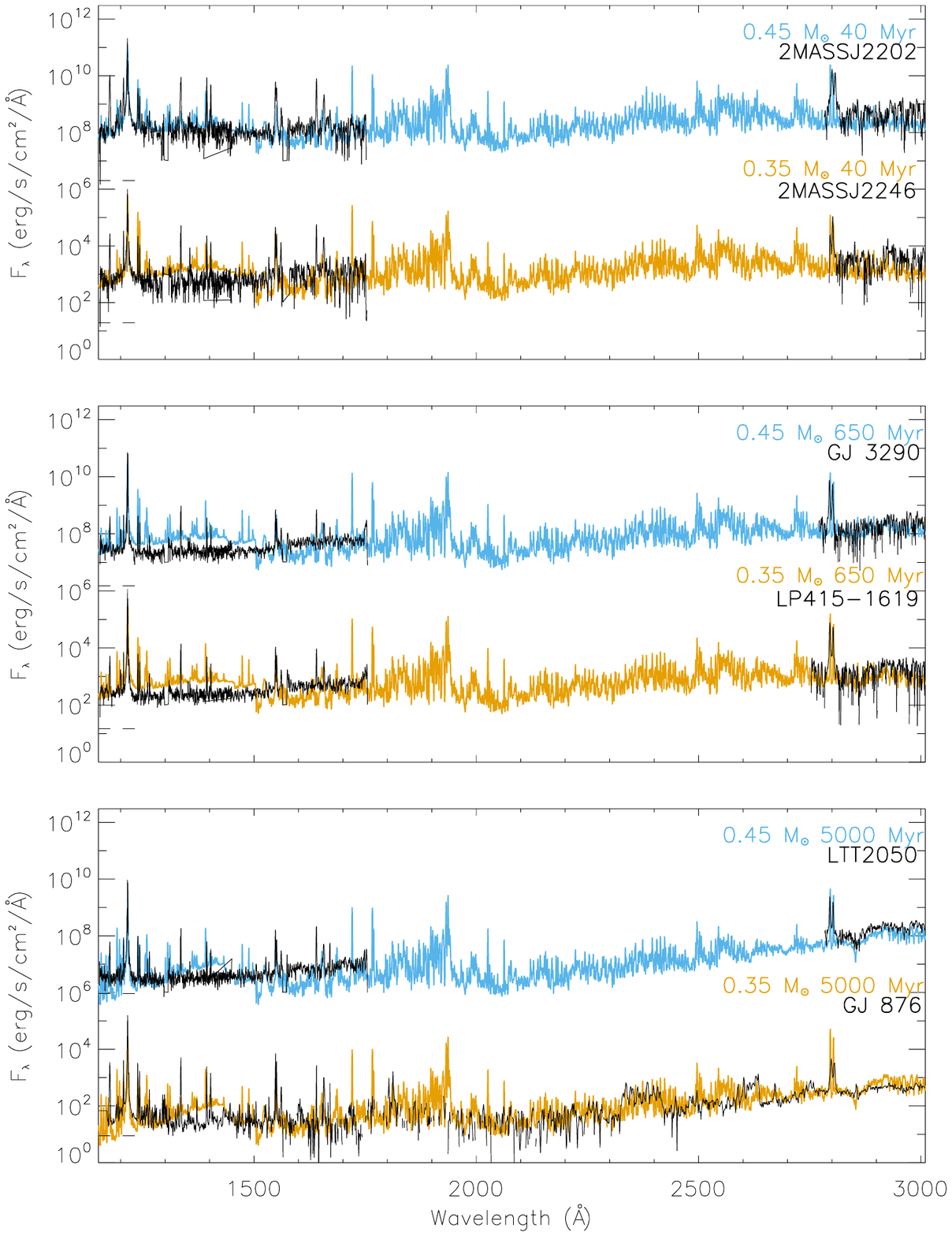}
    \begin{tabular}{lccccccc}
    \hline \hline
    Target & Group & Age & Spectral Type & $T_{\rm eff}$ & Radius & Distance & Radial Velocity \\
           &       & (Myr)&             &   (K)     & (R$_\sun$) & (pc) & (km s$^{-1}$)\\
    \hline
    2MASSJ22025453-6440441\dotfill & Tuc-Hor & 45       & M1.8\tablenotemark{1}& 3802\tablenotemark{5} & 0.74\tablenotemark{6} & 43.7$\pm$0.1\tablenotemark{5} & 0.38$\pm$2.99\tablenotemark{5}\\
    2MASSJ22463471-7353504\dotfill & Tuc-Hor & 45       & M2.3\tablenotemark{1} & 3816\tablenotemark{5} & 0.74\tablenotemark{6} & 50.2$\pm$0.1\tablenotemark{5} & 5.05$\pm$2.18\tablenotemark{5}\\
    GJ 3290\dotfill                 & Hyades  & 650      & M1.5\tablenotemark{2} & 4046\tablenotemark{5} & 0.6\tablenotemark{6}  &46.0$\pm$0.1\tablenotemark{5} & 40.4$\pm$0.4\tablenotemark{5}\\
    LP415-1619\dotfill              & Hyades & 650      & M2.5\tablenotemark{2} & 4111\tablenotemark{5} & 0.6\tablenotemark{6}  & 42.1$\pm$0.1\tablenotemark{5} & -6.7$\pm$0.1\tablenotemark{7}\\
    LTT2050\dotfill                 & Field & $>$2000  & M1.0\tablenotemark{3}  & 3914\tablenotemark{5} & 0.5\tablenotemark{6} & 11.21$\pm$0.01\tablenotemark{5} & -7.13$\pm$0.23\tablenotemark{5}\\
    GJ 876\dotfill                  & Field   & $>$2000  & M4.0\tablenotemark{4} & 3837\tablenotemark{5} & 0.37\tablenotemark{4} & 4.67$\pm$0.01\tablenotemark{5} & -1.52$\pm$0.16\tablenotemark{8}\\
    \hline
    \end{tabular}
        \caption{Median 0.45 $M_\sun$ (blue) and 0.35 $M_\sun$ (orange) 45 Myr, 650 Myr, and field age star synthetic spectra compared to HAZMAT \textit{HST} spectra of similar type and age (black). The GJ 876 spectrum is from the MUSCLES Treasury Survey \citep{France2016}. Bottom Table: Stellar parameters.}
        \tablerefs{(1) \citealt{Kraus2014}, (2) \citealt{Alonso2015}, (3) \citealt{Stephenson1986}, (4) \citealt{vonBraun2014}, (5) \citealt{Gaia2018}, (6) \citealt{Baraffe2015}, (7) \citealt{Nidever2002}, (8) \citealt{Terrien2015}}
        \label{fig:HST}
\end{figure*}

    


Using a $\chi^2$ test, we identify models that best match the FUV-NUV pair of minimum, lower quartile, median, upper quartile, and maximum fractional flux densities per age from the \textit{GALEX} sample (Table \ref{tab:fluxes}, Figure \ref{fig:fivespec_m1}). In some cases, multiple models match a particular statistical quantity, e.g. the maximum $F_{\rm FUV}$/$F_{\rm J}$, but the models do not simultaneously match the partner value, e.g. the maxmimum $F_{\rm NUV}$/$F_{\rm J}$ (Figure \ref{fig:minmax}). Since the \textit{GALEX} FUV and NUV statistical quantities are calculated in isolation, the paired values do not necessarily represent a real object. From the calculated series of models of 10 and 45 Myr stars, there are no models that have a good match to both components of the paired maxima. In these cases, we identify the model that most closely matches the maximum FUV flux density and its real NUV pair. The models that match the real object with the maximum $F_{\rm FUV}$/$F_{\rm J}$ have a higher $F_{\rm EUV}$/$F_{\rm J}$ than those that match the real object with the maximum $F_{\rm NUV}$/$F_{\rm J}$, providing a more comprehensive range of predicted EUV flux densities.

Since the median \textit{GALEX} FUV and NUV flux densities do not necessarily come from the same object, we verify that each median model per age and mass does match a pair of FUV and NUV flux densities from a single object. We further validate that the computed spectra representative of the median 45, 650 Myr, and 5 Gyr stars are in close agreement with \textit{HST} COS spectra from the HAZMAT or MUSCLES \textit{HST} surveys. Raw COS spectra are coadded and corrected for airglow contamination of the \ion{H}{1} Ly$\alpha$ line using the \cite{Bourrier2018} community template. We scale each \textit{HST} spectrum to the surface of the star using a scale factor of $d^2/R_{\star}^2$ and compare them to the median models of corresponding age. We identify the star that most closely matches the median model, per age and mass, by comparing the observed UV continuum and high signal-to-noise emission lines to the synthetic spectra. 

Within the sample of HAZMAT \textit{HST} COS spectra, we find objects that have very similar UV spectra to each of our median 45, 650 Myr models and the 0.45 $M_\sun$ 5 Gyr model. The 5 Gyr 0.35 $M_\sun$ model is most similar to the observed GJ 876 spectrum from \cite{France2016}. The closest matching spectra are shown in Figure \ref{fig:HST}, with the stellar parameters for these objects given in the table. We confirm that the UV continuum slope of each synthetic spectrum is consistent with a real object and find that there is strong general agreement with the observed bright emission lines, with particularly good agreement with the \ion{Mg}{2} \textit{h} and \textit{k} lines and the wings of Ly$\alpha$.

\begin{figure}[t!]
    \centering
    \includegraphics[width=1.0\linewidth]{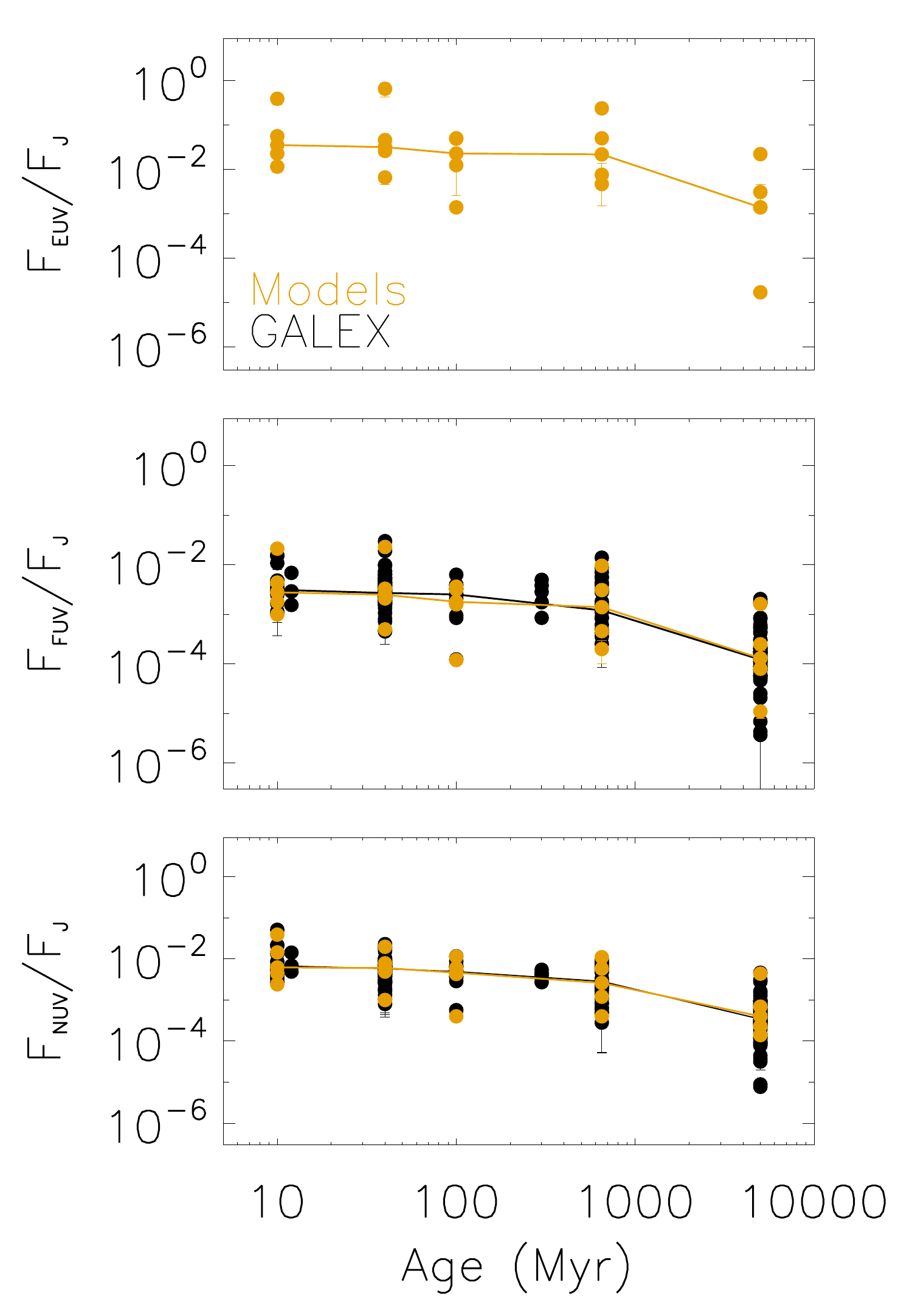}
    \caption{EUV (top), FUV (middle), and NUV (bottom) fractional flux densities as a function of stellar age. Values from the sets (minimum, lower quartile, median, upper quartile, and maximum) of 0.4 $\pm$ 0.05 M$_{\sun}$ models are plotted as orange dots. The median flux densities are connected with an orange line. The \textit{GALEX} FUV and NUV fractional flux densities are plotted as black circles, with the median values connected with a black line.}
    \label{fig:excess3}
\end{figure}

\begin{figure}[th!]
    \centering
    \includegraphics[width=1.0\linewidth]{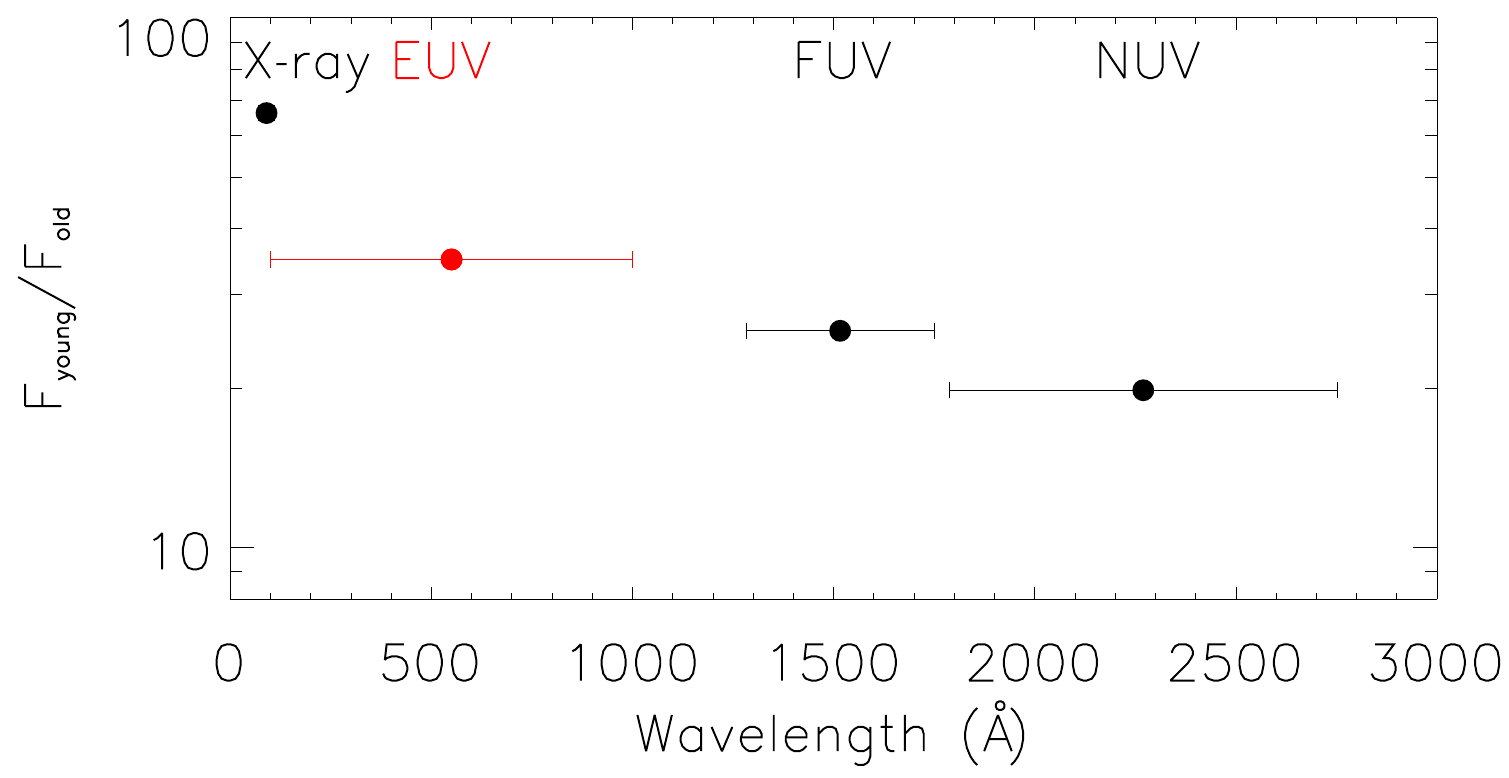}
    \caption{Ratio of measured X-ray, FUV, and NUV (black) fluxes and computed EUV (red) fluxes for M stars in their youth (10 Myr) to field age ($\sim$5 Gyr) decreases monotonically with wavelength.}
    \label{fig:monoton}
\end{figure}

\section{Results}\label{sec:results}

The fractional flux densities for the 0.4 $\pm$ 0.05 $M_\sun$ models (taken as the average of the 0.45 $M_\sun$ and 0.35 $M_\sun$ models) that match the minimum, lower quartile, median, upper quartile, and maximum fractional \textit{GALEX} FUV and NUV flux densities are given in Table \ref{tab:fluxes}. Similar to the observed range in FUV and NUV activity, the modeled EUV fluxes span 1 -- 2 orders of magnitude at each age, shown in Figure \ref{fig:excess3}. We find that the EUV fluxes from the 0.45 $M_\sun$ and 0.35 $M_\sun$ models are similar at each particular age, suggesting that the stellar parameters that spur differences in photospheric flux ($T_{\rm eff}$, $g$, M$_{\star}$) do not greatly impact the upper atmospheric layers from which UV photons are emitted.

\begin{figure}[t!]
    \centering
    \includegraphics[width=1.0\linewidth]{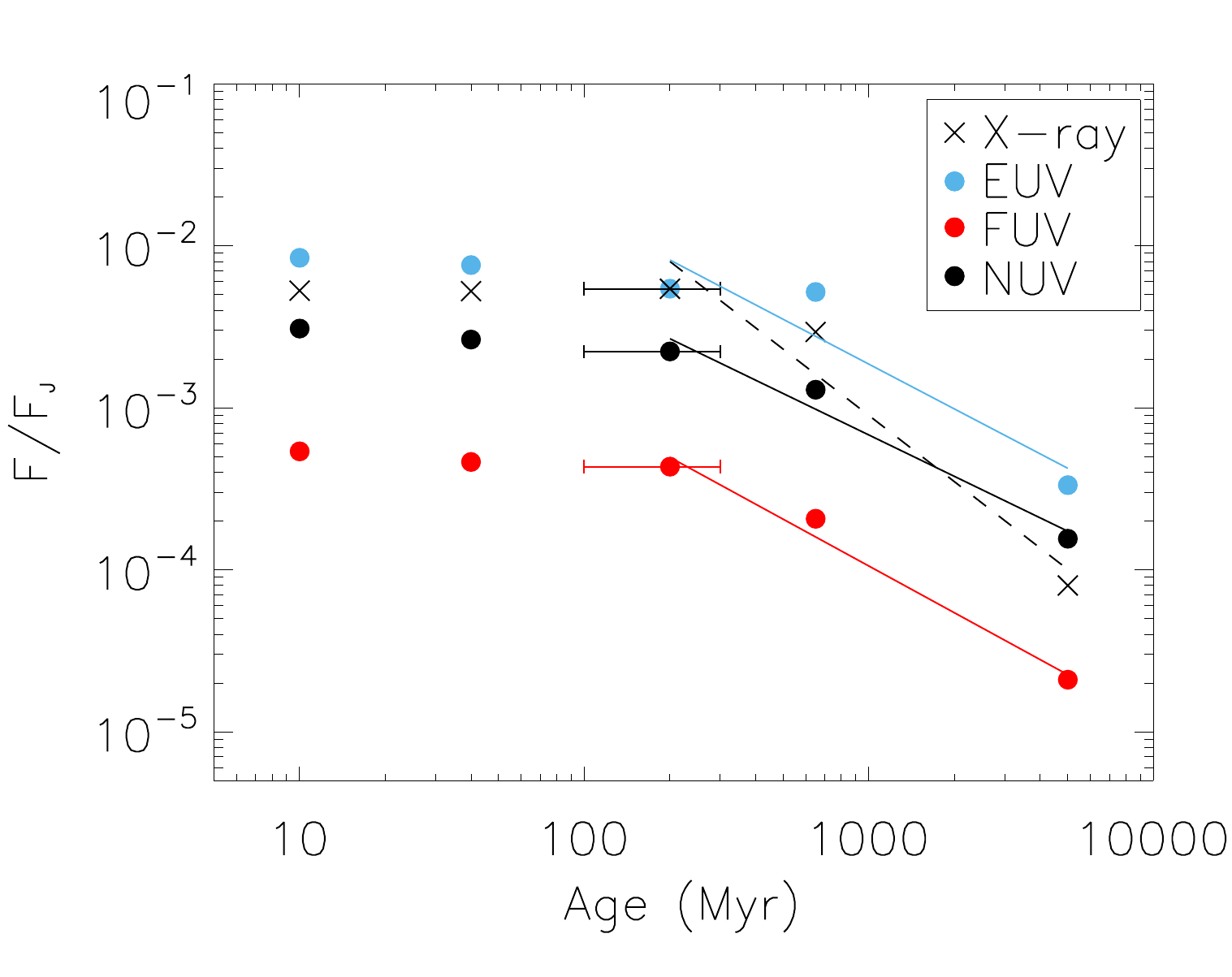}
    \caption{Median X-ray (5--124 \AA), EUV (100--1000 \AA), FUV (1340--1810 \AA), and NUV (1687--3010 \AA) fractional fluxes as a function of stellar age. Median EUV fractional fluxes calculated as the average of the median of 0.45 M$_{\sun}$ and 0.35 M$_{\sun}$ models at each age. Median FUV and NUV fractional flux densities are calculated from the HAZMAT I \textit{GALEX} sample and include upper limits estimated as $(F_{\rm NUV} / F_{\rm J})^{1.11}$. X-ray values are calculated by \cite{Shkolnik2014} from \textit{ROSAT} observations. The coefficients of the power-law fits are given in Table \ref{tab:powerlaw}.}
    \label{fig:excessmed}
\end{figure}

In Figure \ref{fig:excess3}, we plot the excess UV flux densities of the 0.4 $\pm$ 0.05 $M_\sun$ models in orange compared to the \textit{GALEX} measurements in black. The models replicate the spread of \textit{GALEX} FUV and NUV values at each epoch and indicate a similar decline in EUV activity beginning after 120 Myr, with a sharp drop from 650 Myr to field age. The decline in excess EUV flux density falls as $t^{-0.92}$, similar to the drop in measured UV and X-ray flux densities (NUV $\propto t^{-0.85}$, FUV $\propto t^{-0.96}$, X-ray $\propto t^{-1.36}$). We summarize the results from the regression analyses in Table \ref{tab:powerlaw}, given as power laws in the form of $y = \alpha x^\beta$.

Between 10 Myr and 5 Gyr, the median fractional EUV flux decreases by a factor of 35. This is consistent with the observed trend of fractional X-ray, FUV, and NUV fluxes reducing monotonically with wavelength (by approximate factors of 65, 30, and 20, respectively; Figure \ref{fig:monoton}). We show the evolution of median fluxes in Figure \ref{fig:excessmed}, for which the X-ray, EUV, FUV, and NUV flux all remain elevated until a few hundred millions of years followed by a reduction in emission that declines approximately as $t^{-1}$.

The levels of fractional X-ray, FUV, and NUV flux are nearly constant from 10 and 120 Myr, while the median model EUV fractional flux density drops slightly between 45 and 120 Myr and remains constant between 120 and 650 Myr. This mirrors the modeled reproduction of the median FUV flux falling slightly below the observed value for 120 Myr (middle panel of Figure \ref{fig:excess3}) and is related to the simplified linear parameterization used in the construction of the upper atmosphere models. In Section \ref{subsec:correlations} we show that there is a tight correlation between EUV and FUV flux. Using the derived relationship in Table \ref{tab:powerlaw} ($F_{\rm EUV}$ $\propto$ $F_{\rm FUV}$ $^{1.05}$), the factor of 1.4 difference between the median model $F_{\rm FUV}$ and the observed value for 120 Myr M stars suggests that $F_{\rm EUV}$ is underpredicted by a factor of 1.44 at this age. Adjusting the median 120 Myr EUV flux and recalculating the power-law fit yields a relation of $t^{-1.02}$ with an $R^2$ value of 0.94.

\begin{table*}[t!]
    \begin{center}
    \begin{tabular}{lccccr}
    \hline \hline
    $x$ & $y$ & Subset & $\alpha$ & $\beta$ & $R^2$\\
    \hline
    Age & $F_{X}$/$F_{\rm J}$ & Median \textit{ROSAT} $\geq$120 Myr &10.78 & -1.36 & 0.95\\
    Age & $F_{\rm EUV}$/$F_{\rm J}$ & Median Models $\geq$120 Myr& 1.07 (2.38)\tablenotemark{a} & -0.92 (-1.0)\tablenotemark{a} & 0.88 (0.94)\tablenotemark{a}\\
    Age & $F_{\rm FUV}$/$F_{\rm J}$ & Median \textit{GALEX} $\geq$120 Myr & 0.08 & -0.96 & 0.98\\
    Age & $F_{\rm NUV}$/$F_{\rm J}$ & Median \textit{GALEX} $\geq$120 Myr & 0.24 & -0.85 & 0.97\\
    \hline
    $F_{Ly\alpha}$ & $F_{\rm EUV}$ & Median Models & 21.94 & 0.85 & 0.95 \\
    $F_{Ly\alpha}$\tablenotemark{b} & $F_{\rm EUV}$ & Median Models & 379,441 & 1.86 & 0.36 \\
    $F_{\rm FUV}$ & $F_{\rm EUV}$ & Median Models& 18.20 & 1.05 & 0.99\\
    $F_{\rm NUV}$ & $F_{\rm EUV}$ & Median Models& 1.93 & 1.19 & 0.96\\
    \hline
    Age & $F_{100-360\AA}$ & Median Models  &1.62 & -0.72 & 0.84\\
    Age & $F_{920-1180\AA}$ & Median Models & 6.49 & -0.72 & 0.88\\
    Age & $F_{1-1180\AA}$ & Median \textit{ROSAT} \& Median Models   & 26.35 & -0.72 & 0.85\\
    \hline
    $F_{NV}$ & $F_{90-360\AA}$ & \textit{EUVE} \& Median Models  & 0.97 & 0.80 & 0.93 \\
    \hline
    \end{tabular}
    \caption{Power-law Coefficients for $y = \alpha x^\beta$. The relationships are calculated with age in units of Myr, and flux in units of erg cm$^{-2}$ s$^{-1}$ and scaled to 1 AU. \label{tab:powerlaw}}
    \end{center}
    \tablenotetext{a}{Parenthetical values are the fit when the median 120 Myr models are adjusted to be consistent with the median 120 Myr \textit{GALEX} FUV.}
    \tablenotetext{b}{Continuum-subtracted Ly$\alpha$ flux.}
     
\end{table*}

\begin{figure}[t!]
    \centering
    \begin{tabular}{lccc}
    \hline \hline
    & log($\nabla T_{\rm TR}$) & log($m_{\rm TR}$) & log($m_{\rm Tmin}$) \\
    & (K dyne$^{-1}$ cm$^{2}$) & (g cm $^{-2}$) & (g cm $^{-2}$)\\
    \hline
    10 Myr\dotfill  &9 & -5.25 $\pm$ 0.25   & -4  \\
    45 Myr\dotfill  &9 & -5.25 $\pm$ 0.25   & -4  \\ 
    120 Myr\dotfill &9 & -5.5 & -4  \\ 
    650 Myr\dotfill &9 & -5.5 & -4.5 \\ 
    5000 Myr\dotfill   &9 & -6.5 & -5 \\ 
    \hline
    \end{tabular}
    \includegraphics[width=\columnwidth]{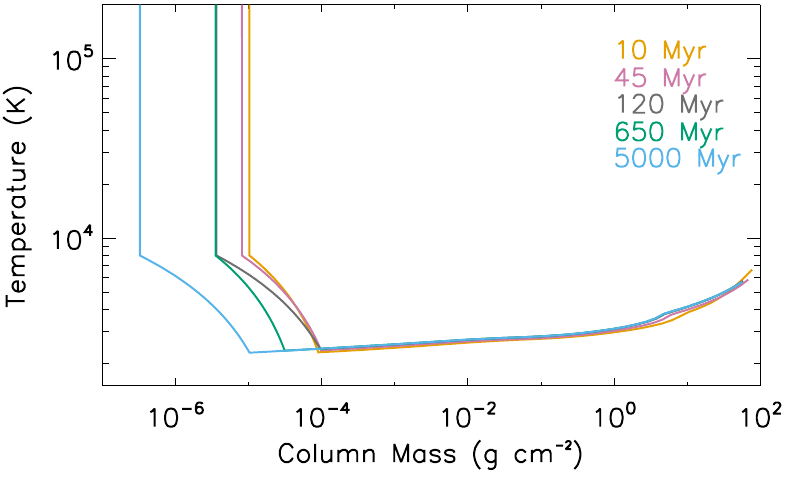}
    \caption{Model parameters (top) and temperature-column mass structures (bottom) for the median models at different ages. At each age, the median models for both the 0.45 $M_\sun$ and 0.35 $M_\sun$ stars are constructed with the same parameter values, with an exception of log($m_{\rm TR}$) at 10 and 45 Myr. The median 0.45 $M_\sun$ models at 10 and 45 Myr are described by log($m_{\rm TR}$) = -5, and the 0.35 $M_\sun$ models are described by log($m_{\rm TR}$) = -5.5.}
    \label{fig:structures}
  \end{figure}

\subsection{Evolution of the Temperature Structure}

As low-mass stars contract onto the main sequence, decreasing in radius and increasing in surface gravity, the chromosphere moves outward and they become less UV active. The temperature structures for the median models are shown in Figure \ref{fig:structures}, along with a table listing the prescribed upper atmospheric model parameters. We find that for both the 0.45 $M_\sun$ and 0.35 $M_\sun$ stars, the median models all have the same temperature gradient in the TR (log($\nabla T_{\rm TR}$) = 9 K dyne$^{-1}$ cm$^{2}$) and the location of the temperature minimum steadily moves toward smaller column mass as the star evolves. 

Between 10 and 45 Myr, the radius and surface gravity of a particular mass star changes, while $T_{\rm eff}$ remains constant (Table \ref{tab:ageparams}). Separately in the 0.45 $M_\sun$ and 0.35 $M_\sun$ models, the observed median UV fluxes between 10 and 45 Myr are generated by temperature structures with the same trio of upper atmospheric model parameters. Since the median $F_{\rm UV}$/$F_{\rm J}$ flux is relatively constant between these ages, it suggests that surface gravity does not have a large influence on the upper atmosphere nor on the emergent UV flux. Comparing the 0.45 $M_\sun$ models to the 0.35 $M_\sun$ models at these ages, which have the same surface gravity but differ by 150 K and 0.1 $M_\sun$, the chromospheric structures that best reproduce the median FUV and NUV excess flux densities are very similar, with a difference in log($m_{\rm TR}$) of just 0.5 g cm$^{-2}$. Since the EUV spectrum is very sensitive to changes in $m_{\rm TR}$, this deviation results in a factor of two difference in fractional $F_{\rm EUV}/F_{\rm J}$ flux between the two models.
 
The median upper atmospheric temperature structures for both the 0.45 $M_\sun$ and 0.35 $M_\sun$ models at ages $\geq$ 120 Myr are described by the same set of parameters, regardless of the differences in $T_{\rm eff}$ and log(\textit{g}). The resulting median EUV fluxes at each age $\geq$ 120 Myr are the same between the 0.45 $M_\sun$ and 0.35 $M_\sun$ models.

\cite{Parsons2018} showed that the \cite{Baraffe2015} evolutionary models underpredict M dwarf radii by $\sim$6\% and overpredict the effective temperature by $\sim$100 K. Our models suggest that differences on this scale for $T_{\rm eff}$, log(\textit{g}), and mass do not affect the atmosphere above the photosphere nor our predicted EUV spectra.

\subsection{Evolution of the Full UV Spectrum}

\begin{turnpage}
\begin{figure*}[h!]
    \centering
    \includegraphics[scale=0.8]{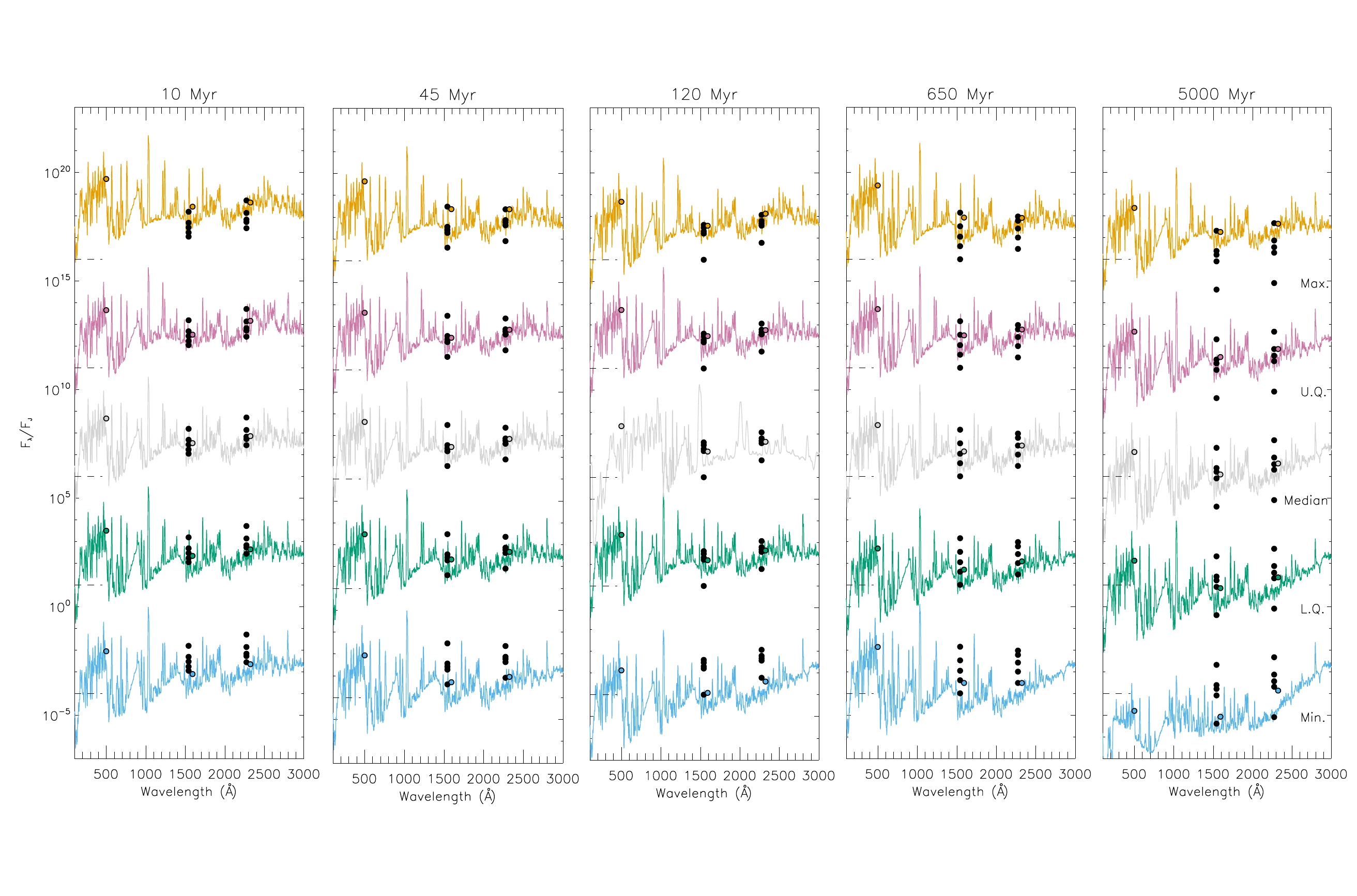}
    \caption{PHOENIX synthetic spectra that match the minimum (blue), lower quartile (green), median (grey), upper quartile (pink), and maximum (yellow) \textit{GALEX} measurements at each age.  Synthetic photometry from the PHOENIX models are plotted as circles in corresponding colors, calculated over $\lambda_{\rm EUV}$ = 100--1170 \AA \ and the same wavelengths as the \textit{GALEX} FUV ($\lambda_{\rm FUV}$ = 1340--1810 \AA) and NUV ($\lambda_{\rm NUV}$ = 1690--3010 \AA) filter profiles. The spectra and photometry have been vertically shifted by 10$^4$ for clarity and the wavelength resolution has been degraded for clarity. Minimum, lower quartile, median, upper quartile, and maximum \textit{GALEX} photometric points are plotted as black circles. The minima for the 45 and 650 Myr stars are upper limits.}
    \label{fig:fivespec_m1}
\end{figure*}
\end{turnpage}

We present the synthetic UV spectra of the 0.4 $\pm$ 0.05 $M_\sun$ models that match the minimum, lower quartile, median, upper quartile, and maximum \textit{GALEX} measurements per age in Figure \ref{fig:fivespec_m1}. The full high-resolution spectra (100 \AA -- 5.5 $\mu$m, $\Delta\lambda\sim$0.01 \AA) are available at MAST via
\dataset[https://doi.org/10.17909//t9-j6bz-5g89]{https://doi.org/10.17909//t9-j6bz-5g89}.\footnote{\url{https://archive.stsci.edu/hlsp/hazmat}}

\begin{figure*}[t!]
    \centering
    \includegraphics[width=1.0\textwidth]{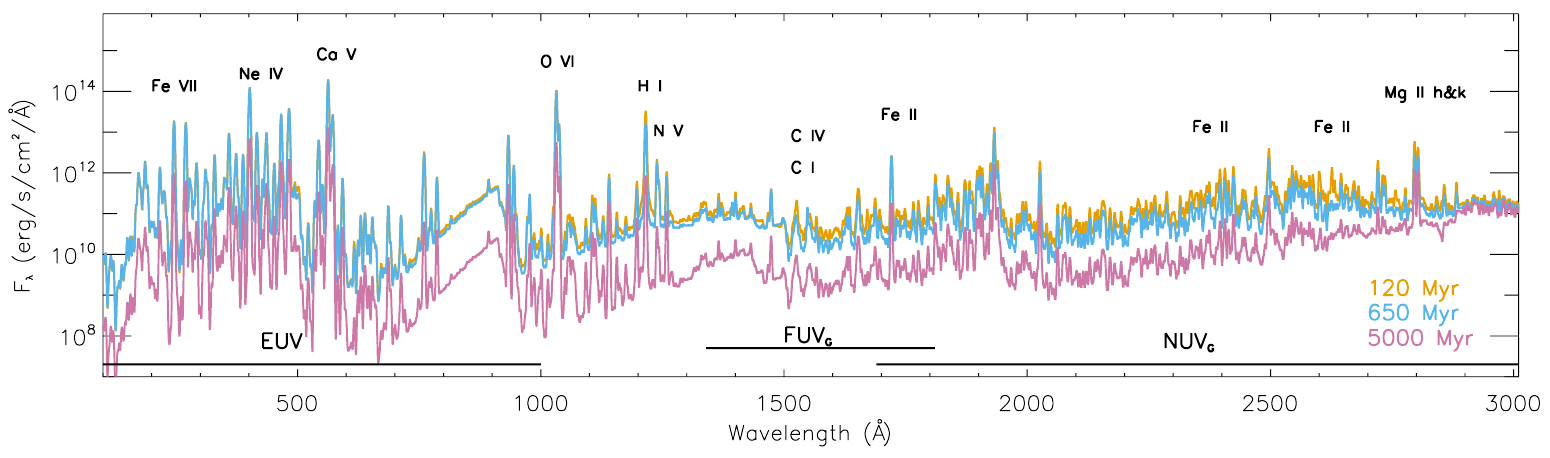}
    \caption{Synthetic UV spectra of median 0.4 $\pm$ 0.05 $M_\sun$ models at 120 Myr (orange), 650 Myr (blue), and 5 Gyr (purple). Prominent emission features and the wavelength ranges used to compute band-integrated fluxes are labeled in the top panel. Spectral resolution has been degraded for clarity. Some of the EUV emission lines with higher ionization stages are currently computed in local thermodynamic equilibrium and are likely overpredicted; for example, the strong \ion{O}{6} lines near 1000 \AA \ are likely overpredicted by up to a factor of 10.}
    \label{fig:uvevol}
\end{figure*}

\begin{figure*}[t!]
    \centering
    \includegraphics[width=1.0\textwidth]{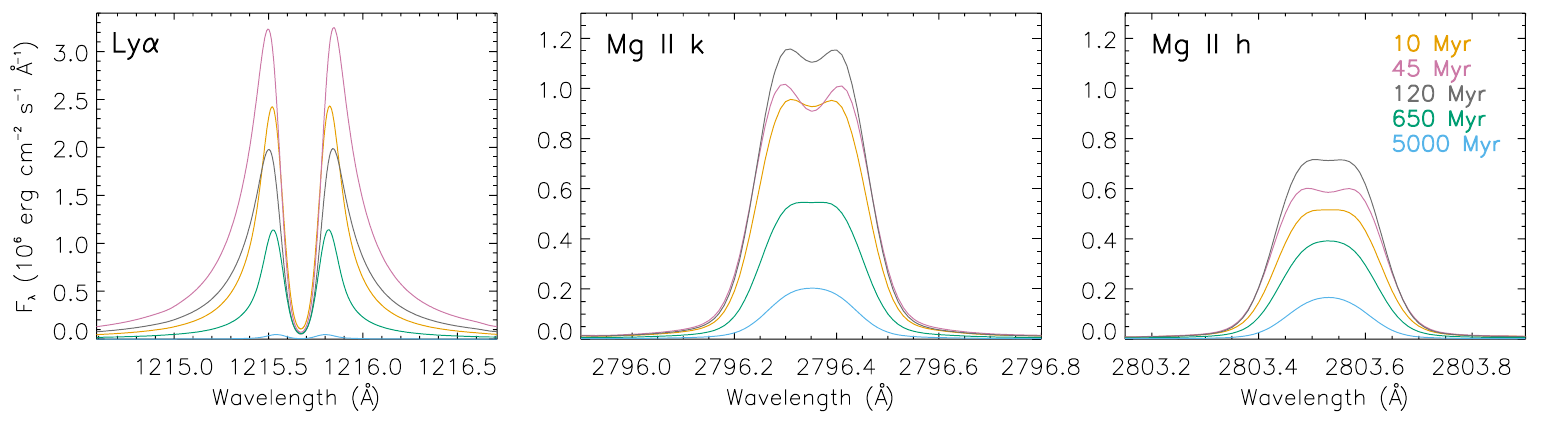}
    \includegraphics[width=1.0\textwidth]{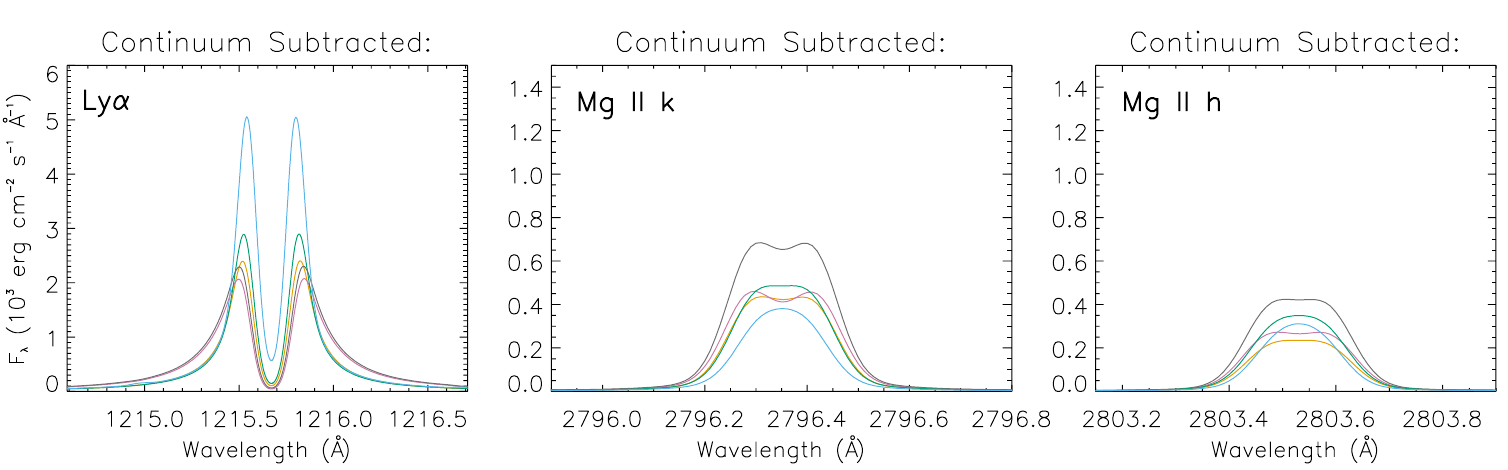}
    \caption{Computed line profiles of \ion{H}{1} Ly$\alpha$ and \ion{Mg}{2} h \& k for the median models per age. Top: line profiles including the contribution from the continuum. We note that, on this scale, the 5000 Myr Ly$\alpha$ line profile is along the X-axis. Bottom: Continuum-subtracted profiles.}
    \label{fig:lyamg}
\end{figure*}

To illustrate the differences between the UV spectra at each age, we compare the median models for 120, 650 Myr, and 5 Gyr in Figure \ref{fig:uvevol}. The 10 and 45 Myr spectra are excluded for clarity. The UV spectra are similar for all models $\leq$ 650 Myr with the UV pseudocontinuum decreasing subtly with age. There are also subtle differences in the slope of the pseudocontinuum surrounding the Ly$\alpha$ line from 1000 to 1500 \AA \ and in the strength of the \ion{Fe}{2} lines from 2300 to 2700 \AA. The computed EUV spectrum is sensitive to the prescription of $\nabla T_{\rm TR}$ and $m_{\rm TR}$. Since these parameters remain nearly constant between 10 and 650 Myr, the EUV spectra are very similar. The decrease in log($m_{\rm TR}$) of $\sim$ 0.25 g cm$^{-2}$ between the 10 -- 45 Myr stars and 120 -- 650 Myr stars results in a slight decrease in the EUV continuum; the integrated EUV flux decreases by a factor of 1.5. The larger deviation in the model parameters of the 5 Gyr stars causes a more noticeable decrease in UV continuum flux levels and an increase in slope, as shown in Figure \ref{fig:uvevol}. 

\begin{figure*}[th!]
    \centering
    \includegraphics[width=1.0\textwidth]{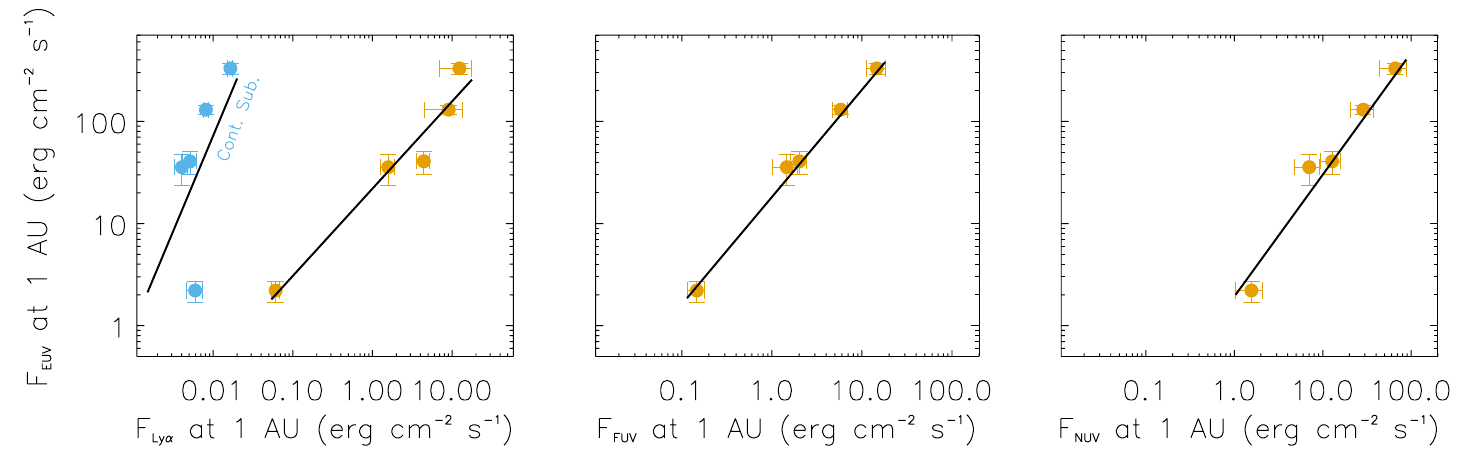}
    \caption{Integrated EUV (100 -- 1000 \AA) flux versus integrated Ly$\alpha$ (1211.5 -- 1219.7 \AA), FUV (1340 -- 1810 \AA), and NUV (1690 -- 3010 \AA) fluxes scaled to 1 AU for median 0.4 $\pm$ 0.05 $M_\sun$ models at each age. The coefficients of the power-law fits are given in Table \ref{tab:powerlaw}. Left panel: Continuum-subtracted Ly$\alpha$ line fluxes are plotted in blue.}
    \label{fig:elfn}
\end{figure*}

The evolution of individual line fluxes for \ion{H}{1} Ly$\alpha$ and \ion{Mg}{2} \textit{h} and \textit{k} are shown in Figure \ref{fig:lyamg}, with the continuum-subtracted line profiles shown in the bottom panel. When not continuum-subtracted, the strength of the Ly$\alpha$ line generally decreases with age. There is an exception, however, with the strength of the 45 Myr model Ly$\alpha$ profile exceeding the 10 Myr profile. This is due to a difference in log($g$) of $\sim$ 0.4 cm s$^{-2}$, with the larger gravity yielding 1.7 times more Ly$\alpha$ emission. When the Ly$\alpha$ line profiles are continuum-subtracted, there is no uniform decrease of line flux with age, but rather near-constant profiles from 10 to 650 Myr, followed by an increase in line flux relative to the surrounding pseudocontinuum for the 5 Gyr models. This is tied to the location of the TR; the pseudocontinuum surrounding Ly$\alpha$ is sensitive to the \textit{m$_{\rm TR}$} parameter, shifting downwards by a factor of five as \textit{m$_{\rm TR}$} is decreased by 0.5 g cm$^{-2}$.

There is a different trend observed in the strength of the \ion{Mg}{2} doublets, where the line profiles increase with age from 10 to 120 Myr, and then decrease toward field age. Relative to the continuum, the \ion{Mg}{2} profiles are fairly consistent, but with a  marginally stronger \ion{Mg}{2} k line in the 120 Myr model compared to the other ages. These lines form near the top of the chromosphere/bottom of the TR ($\sim$10$^{4.2}$~K, \citealt{Sim2005}). The differences in the line profiles illustrate the sensitivity of these lines to the temperature gradient in the chromosphere in combination with the location of the TR. Further discrepancies also suggest that the photospheric parameters ($T_{\rm eff}$, $g$) influence the NUV spectrum at longer wavelengths.

Intrinsic Ly$\alpha$ profiles for M stars are not well understood because the observed profiles are contaminated by interstellar hydrogen absorbing nearly the entire Ly$\alpha$ core. Reconstructions must even be preformed for M stars with high ($>$ 150 km s$^{-1}$) radial velocities such that the Ly$\alpha$ line is Doppler shifted enough to observe the majority of the intrinsic profile \citep{Guinan2016,Schneider2019}. Since the observed line profiles of Ly$\alpha$ and \ion{Mg}{2} \textit{h} and \textit{k} are similar in the Sun \citep{Donnelly1994,Lemaire1998}, groups such as \cite{Wood2005} and \cite{Youngblood16} have estimated the shape of the central portion of Ly$\alpha$ profiles for several M stars using \ion{Mg}{2} \textit{h} and \textit{k} observations. The \ion{Mg}{2} lines, however, form at slightly lower temperatures than Ly$\alpha$, and our computed profiles show that the depth and/or occurrence of a self-reversal\footnote{Optically thick resonance lines like Ly$\alpha$ and \ion{Mg}{2} \textit{h} and \textit{k} may present with self-reversals as a result of the line wings forming deeper in the atmosphere, where the source function is increasing with temperature and the atmosphere is relatively close to thermal equilibrium, while the core forms in the upper atmosphere, where departures from LTE are large and emerging photons are no longer coupled to the local temperature.} in the \ion{Mg}{2} \textit{h} and \textit{k} lines does not correspond to the shape of the Ly$\alpha$ core. For example, Figure \ref{fig:lyamg} shows that at field ages, the \ion{Mg}{2} \textit{h} and \textit{k} profiles present as single peaked, while the Ly$\alpha$ profile maintains a self-reversal. Further, our models indicate that over a stellar lifetime, the line profiles do not evolve in similar ways. Due to these differences, we suggest further investigation regarding the reliability of using observed \ion{Mg}{2} \textit{h} and \textit{k} line profiles to estimate the core of Ly$\alpha$.

\subsection{Integrated Flux Correlations}\label{subsec:correlations}

\begin{figure*}[t!]
    \centering
    \includegraphics[width=1.0\textwidth]{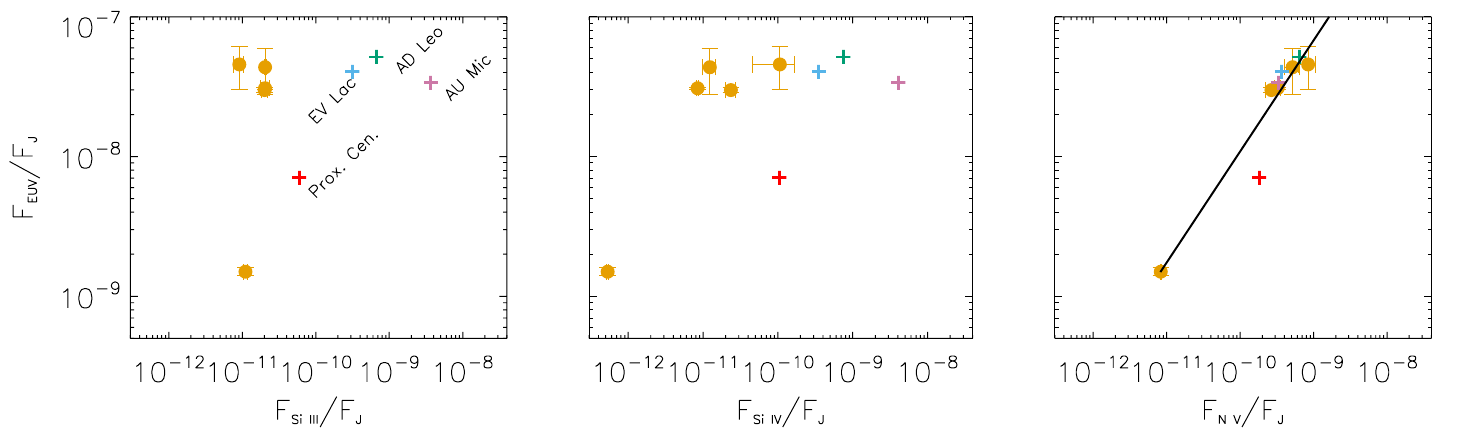}
    \caption{Fractional $F_{\rm EUV}/F_{\rm J}$ fluxes (90 -- 360 \AA) versus fractional \ion{Si}{3}, \ion{Si}{4}, and \ion{N}{5} fluxes for median 0.4 $\pm$ 0.05 $M_\sun$ models at each age (filled circles) compared to observed FUV line fluxes of select M stars from \cite{France2018} (crosses). Fluxes for the models and observed stars are listed in Table \ref{tab:trflux}. The fractional EUV flux and \ion{N}{5} flux of the combined set of observations and models can be fit with the line: $0.97 x^{0.8}$ ($R^2$ = 0.93).}
    \label{fig:euvfuvlines}
\end{figure*}

\begin{table*}[th!]
    \centering
    \begin{tabular}{lcccc}
    \hline \hline
          & $F_{90-360 \AA}$/$F_{\rm J}$ & $F_{SiIII}$/$F_{\rm J}$ & $F_{SiIV}$/$F_{\rm J}$ & $F_{NV}$/$F_{\rm J}$\\
          & ($\times$ 10$^{-8}$) & ($\times$ 10$^{-11}$)& ($\times$ 10$^{-11}$)& ($\times$ 10$^{-11}$)\\
 \hline
10 Myr	&4.57	$\pm$	1.55	&	0.91	$\pm$	0.14	&	10.62	$\pm$	6.06	&	84.09	$\pm$	22.63	\\
45 Myr	&4.37	$\pm$	1.57	&	2.05	$\pm$	0.06	&	1.23	$\pm$	0.25	&	51.75	$\pm$	11.88	\\
120 Myr	&3.09	$\pm$	0.06	&	2.04	$\pm$	0.27	&	0.85	$\pm$	0.04	&	32.35	$\pm$	3.10	\\
650 Myr	&3.00	$\pm$	0.12	&	2.01	$\pm$	0.20	&	2.35	$\pm$	0.37	&	26.67	$\pm$	4.23	\\
5000 Myr	&0.15	$\pm$	0.01	&	1.11	$\pm$	0.06	&	0.06	$\pm$	0.01	&	0.83	$\pm$	0.01	\\
    \hline
    AU Mic\tablenotemark{1,2} &       5.16& 66.67 $\pm$ 0.94& 	74.65 $\pm$ 0.47& 63.85 $\pm$ 	0.47		\\
    AD Leo\tablenotemark{1,2} &       3.38& 369.95$\pm$ 0.47&    405.63$\pm$ 0.47& 32.86 $\pm$ 	0.47		\\
    EV Lac\tablenotemark{1,2} &       4.05& 31.53 $\pm$ 1.81&    35.14 $\pm$ 0.90& 36.94$\pm$   0.90      \\
    Proxima Cen\tablenotemark{1,2} & 0.70&  5.92 $\pm$ 0.42&     10.42 $\pm$ 0.23& 18.17$\pm$ 0.28		\\
    \hline
    \end{tabular}
    \caption{Fractional transition region line fluxes for the median models (top) and observations (bottom) plotted in Figure \ref{fig:euvfuvlines}. References for measured fluxes are (1) \cite{France2018} for the integrated EUV (90 -- 360 \AA) flux and FUV line fluxes, and (2) \cite{Zacharias2004} for J band fluxes.}
    \label{tab:trflux}
\end{table*}

Correlations amongst observable stellar activity diagnostics provide important information about line formation mechanisms and have been used as proxies to predict broadband EUV flux estimates in lieu of available observations (See Appendix \ref{app:euvlya}). In Figure \ref{fig:elfn}, we compare the integrated EUV flux of our median 0.4 $\pm$ 0.05 $M_\sun$ models per age to the Ly$\alpha$ line flux and FUV and NUV fluxes, integrated over the same wavelengths as the \textit{GALEX} filter profiles. The fluxes are scaled to 1 AU using a radius-squared relationship with radii obtained from the \cite{Baraffe2015} evolutionary tables (Table \ref{tab:ageparams}). We consider the Ly$\alpha$ line, which forms over extensive depths in the stellar atmosphere ($T_{\rm form,core}$ = $\sim$10$^{4.5}$~K, $T_{\rm form,wings}$ = $\sim$10$^3$ K; \citealt{Sim2005}), since it is the strongest emitting line in the UV spectrum but is outside of the wavelength range covered by the \textit{GALEX} FUV filter.

A strong correlation has been found between surface Ly$\alpha$ and FUV excess flux for M and K stars ($R^2$=0.91) \citep{Shkolnik2014b}; we find that Ly$\alpha$ similarly has a strong correlation with the total integrated EUV flux ($R^2$ = 0.95, Table \ref{tab:powerlaw}). Relative to the surrounding pseudocontinuum, however, the outlying 5 Gyr models drive a poor power-law fit ($R^2$ = 0.36), since these models have the lowest EUV flux, but largest continuum-subtracted Ly$\alpha$ flux. Allowing for the possibility that this could be a relic of our simplified temperature structure, we calculate an additional fit to the continuum-subtracted $F_{\rm Ly\alpha}$ versus $F_{\rm EUV}$ excluding the 5 Gyr models, and find an improved relationship $\propto x^{1.46}$ ($R^2$ = 0.9).

Correlations have been found among simultaneous FUV and NUV observations \citep{Shkolnik2014, Miles2017} and simultaneous X-ray and UV observations \citep{Mitra2005} of M dwarfs, suggesting that the layers in stellar upper atmospheres (chromosphere, TR, and corona) are heated by similar processes. We find similar correlations between $F_{\rm EUV}$ and both $F_{\rm FUV}$ and $F_{\rm NUV}$ across all ages for our median models in Figure \ref{fig:elfn}, with $R^2$ values of 0.99 and 0.96. The power-law coefficients describing these correlations are given in Table \ref{tab:powerlaw}. The EUV and FUV flux decrease together at nearly the same rate and indicates that the EUV and FUV spectrum form over the same range of temperatures in the stellar atmosphere.

\begin{figure*}[t!]
    \centering
    \includegraphics[width=1.0\textwidth]{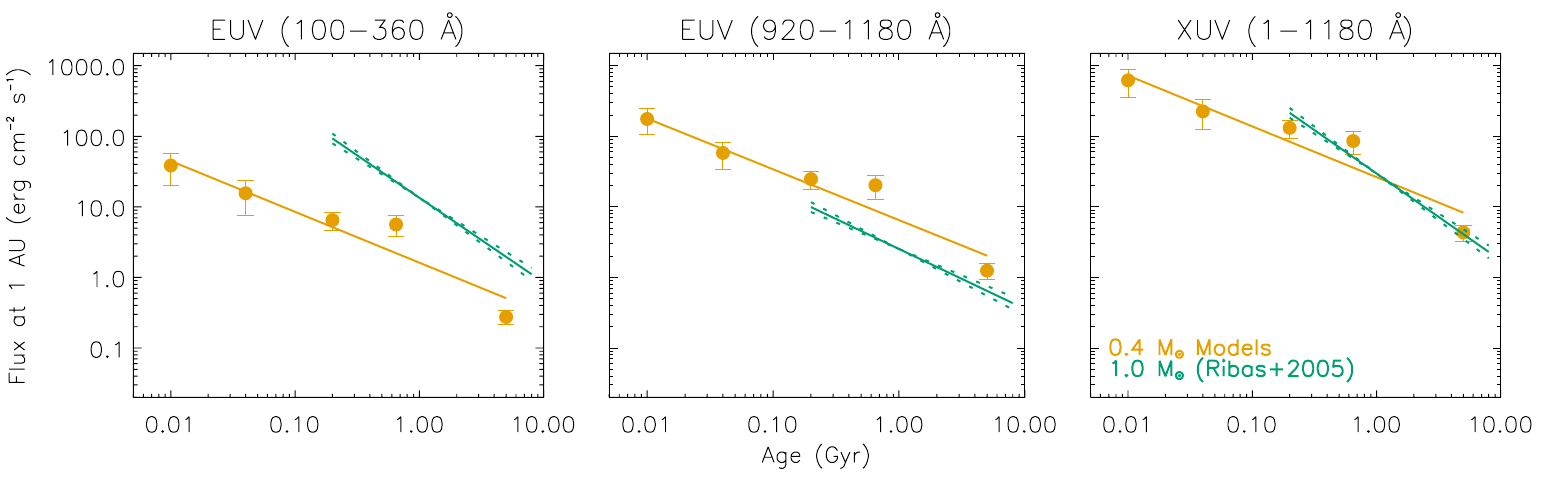}
    \caption{Integrated fluxes scaled to 1 AU versus age for median 0.4 $\pm$ 0.05 $M_\sun$ models. Power-law fits to the models are plotted as orange lines. Power-law fits to observations of Sun-like stars from \cite{Ribas2005} are plotted in green. The coefficients of the power-law fits to the PHOENIX models are given in Table \ref{tab:powerlaw}.}
    \label{fig:sunintime}
\end{figure*}

\cite{France2018} looked for correlations between EUV flux (90 -- 360 \AA) and TR lines in FUV \textit{HST} observations of FGKM stars using a combination of archival \textit{EUVE} observations and model EUV predictions. We present a similar comparison in Figure \ref{fig:euvfuvlines} of fractional EUV flux versus \ion{Si}{3} (1206.49 \AA, 1206.55 \AA), \ion{Si}{4} (1393.75 \AA, 1402.76 \AA), and \ion{N}{5} (1238.82 \AA, 1242.8 \AA). The FUV lines have formation temperatures at different depths in the TR: \ion{Si}{3} (10$^{4.5}$~K), \ion{Si}{4} (10$^{4.8}$--10$^{5.1}$~K), \ion{N}{5} (10$^{5.2}$~K) \citep{Sim2005}. We plot our median models along with the M stars in \cite{France2018} that have interstellar medium (ISM)-corrected \textit{EUVE} fluxes and \ion{Si}{3}, \ion{Si}{4}, and \ion{N}{5} line fluxes available. The observed sample consists of active stars of various ages: AU Mic (24 Myr; \citealt{Bell2015}), AD Leo (25--300 Myr; \citealt{Shkolnik2009}), EV Lac (45 Myr; \citealt{Parsamyan1995}), and Proxima Centauri ($\sim$6 Gyr; \citealt{Morel2018}). We compute the fractional fluxes using \textit{J}$_{\rm 2MASS}$ magnitudes from \cite{Skrutskie2006}.

Integrating over 90 -- 360 \AA, we find that our median models produce similar levels of fractional EUV flux as the observed \textit{EUVE} spectra, with particularly good agreement amongst the three young stars (Figure \ref{fig:euvfuvlines}, Table \ref{tab:trflux}). The fractional EUV flux for Proxima Centauri is higher than the median 5 Gyr model, falling within the values of our upper quartile and maximum models for the age ($F_{\rm EUV}$/$F_{\rm J}$ = (0.5--3) $\times$ 10$^{-8}$). The fractional EUV flux and FUV line fluxes of the models and observations are positively correlated; however, the models predict lower \ion{Si}{3} and \ion{Si}{4} line fluxes than the observed sample. We note that the EUV and FUV observations are non-contemporaneous which contributes uncertainty and may explain some of the disagreement between our model fluxes and the observations; however, the discrepancies in the Si fluxes are likely due to current shortcomings of the models. The maximum temperature in our thermal structures do not reach coronal levels and we do not include the downward flow of heat and hydrogen ionization from these $\sim$10$^6$ K layers via ambipolar diffusion. These physical processes affect where the onset of the TR begins, near where the Si lines are forming. In future work we will quantify the importance of ambipolar diffusion at these depths. In both data sets, there is a weak correlation between the fractional EUV flux and the Si lines, and a strong correlation between the fractional EUV flux and \ion{N}{5}. These results support the finding of \cite{France2018} that FUV lines with higher formation temperatures are better proxies for fractional EUV flux.

\section{Discussion}\label{sec:discussion}

\subsection{Comparison to the EUV Evolution of G Stars}

The likelihood of M dwarfs hosting habitable worlds has been subject to debate due to major differences between the stellar properties of these cool, yet active, stars compared to hotter and more luminous stars like our Sun. We compare our modeled EUV evolution of early M stars to observations of solar proxies from the Sun in Time program \citep{Ribas2005}. The study considered X-ray and EUV observations of G stars $>$~100~Myr from the \textit{Advanced Satellite for Cosmology and Astrophysics} (\textit{ASCA}, 1 -- 20 \AA), \textit{ROSAT} (20 -- 100 \AA), \textit{EUVE} (100 -- 360 \AA), and \textit{FUSE} (920 -- 1180 \AA). Hindered by the same observational restrictions due to ISM absorption in the 360 -- 912 \AA \ interval, the authors inferred the total integrated flux in this region from interpolations of the flux evolution in the surrounding wavelengths. Due to the large uncertainties associated with this method, in Figure \ref{fig:sunintime} we compare our results to the observed EUV wavelengths (100--360, 920--1180 \AA) separately, in addition to the total integrated XUV (1 -- 1180 \AA) flux.

We find that the modeled fluxes (scaled by ($R_\star$/1 AU)$^2$; $R_\star$ given in Table \ref{tab:ageparams}) show excellent inverse correlation with stellar age, and best-fit with similar power-law relationships as the G stars. However, the slope of the line fit to the fluxes in the 100 -- 360 \AA \ region is more shallow for M stars than the G stars, and is fairly consistent with the slope of the modeled 920 -- 1180 \AA \ interval. While the solar relationships decrease monotonically with each wavelength band from the X-rays to UV, our results indicate that rate of decrease in emission is fairly consistent across EUV through NUV wavelengths for early M stars. This suggests that, in M stars, the plasma in the entire TR cools at nearly the same rate. 

Our models do not include coronal flux contribution and therefore do not predict the X-ray spectrum. To compute the XUV flux, we added the median \textit{ROSAT} X-ray fluxes from \cite{Shkolnik2014} to the median model EUV fluxes. We note that the model+\textit{ROSAT} fluxes cover the wavelength range 20 -- 1180 \AA, with no included contribution from the 1 -- 20 \AA \ region. We find that there is good agreement between the total integrated XUV fluxes of our models and the solar analogues. The overall slope for our models 10 -- 5000 Myr is slightly more shallow than for the G stars. However, when considering the same age range as the Sun in Time program, the $>$ 120 Myr stars show excellent agreement between the two spectral types.

\subsection{Implications for Planet Habitability}

The low luminosities and effective temperatures of M stars means potentially habitable planets reside relatively close to the host star (0.1 -- 0.4 au); however, the location of the canonically defined habitable zone will migrate over time. Over the first hundreds of million years of an M star's lifetime, the shrinking radii and near-constant effective temperatures result in a decrease in luminosity that translates to an inward migration of this temperate region. This means that a planet located in the present-day habitable zone was likely beyond the inner edge in its first few hundred million years. Considering both the elevated bolometric flux of M stars at this time and our finding that $F_{\rm XUV}$/$F_{\rm J}$ is elevated by $\sim$100x relative to field ages, these planets are exposed to $F_{\rm XUV}$ fluxes well over 100 times what they will experience when they are forming their primary and/or secondary atmospheres later in life. The close proximity of the M dwarf habitable zone means that a planet located near the late-stage inner edge would have been exposed to $\sim$100 times more XUV radiation than the Earth received over its lifetime. In its youth, it would have been bombarded with up to 10,000 times more XUV radiation than the Earth receives at present-day. Prolonged exposure of a terrestrial planet to this amount of XUV radiation can lead to thermal escape resulting in the loss of a significant fraction of lighter atmospheric elements and several global oceans' worth of H$_2$O \citep{Luger2015}. 

\subsection{Future Work}

In this study, we considered 0.4 $\pm$ 0.05 $M_\sun$ stars representative of an average early M star and identified the temperature profiles by comparing the emergent spectra to the \textit{GALEX} measurements from the HAZMAT I sample, which consists of early-to-mid M stars ($\geq$ 0.35 $M_\sun$). \cite{Schneider2018} (HAZMAT III) conducted an extension of this study to mid- and late-type M stars (0.08 -- 0.35 $M_\sun$) and found that these fully convective stars do not follow the same UV evolution as the early Ms. Results showed later-type M stars retain higher levels of UV activity out to much older ages. Between the young and field age stars, the NUV and FUV flux density decreases by an average factor of four, compared to a factor of 16 and 24 for early M stars.

\begin{table}[t!]
    \centering
    \begin{tabular}{lcc|cc}
    \hline \hline
   Age & \multicolumn{2}{c}{\textbf{$F_{\rm FUV}$/$F_{\rm J} \times 10^3$}} & \multicolumn{2}{c}{\textbf{$F_{\rm NUV}$/$F_{\rm J} \times 10^3$}}\\
    \hline
    &0.35 M$_{\sun}$ & \textit{GALEX} &  0.35 M$_{\sun}$ & \textit{GALEX}\\
    & Models& Late Ms & Models& Late Ms\\
    \hline
     \textbf{10 Myr}         &2.2&1.5 $^{+1.3}_{-0.5}$    & 5.3&	3.8 $^{+1.6}_{-1.1}$\\
     \textbf{45 Myr}         &2.0&2.6 $^{+0.9}_{-1.5}$  &	5.0&	4.1 $^{+1.8}_{-1.1}$\\
     \textbf{120 Myr}        &1.8&2.1 $^{+2.7}_{-0.6}$  &    4.4&	3.9 $^{+2.3}_{-1.7}$\\
     \textbf{650 Myr}        &1.3&1.7 $^{+1.1}_{-1.3}$  &	  2.5&	3.2 $^{+2.3}_{-1.4}$\\
      \textbf{5 Gyr}         &0.1&0.1 $^{+1.0}_{-1.0}$  &    0.4&	1.1 $^{+1.2}_{-0.9}$\\
    \hline
    \end{tabular}
    \caption{Median fractional flux densities (multiplied by 10$^3$ for clarity) of synthetic spectra representative of 0.35 $M_\sun$ stars at various ages compared to median \textit{GALEX} measurements of late M stars (0.08 -- 0.35 $M_\sun$) from \cite{Schneider2018}. Ranges in the \textit{GALEX} measurements represent inner quartiles of the sample.}
    \label{tab:hz3fluxes}
\end{table}

Considering the HAZMAT III sample, we compare our model estimated median spectra for 0.35 $M_\sun$ stars to the late Ms in \cite{Schneider2018} (Table \ref{tab:hz3fluxes}). We find that the models representative of M1 -- M2 stars are consistent within the inner quartiles of the late Ms in the HAZMAT III sample out to field ages, with marginal deviation at NUV wavelengths. Identifying the models that more closely reproduce the median excess FUV and NUV flux densities from the HAZMAT III sample, the only differences between those in Figure \ref{fig:structures} are shifts in log($m_{\rm Tmin}$) by 0.5 g cm$^{-2}$ for the 10 Myr (log($m_{\rm Tmin}$) = -4.5) and 5 Gyr models (log($m_{\rm Tmin}$) = -4.5). This change does not impact the predicted EUV spectrum, but does imply a difference in the evolution of the temperature structure. From this test, we estimate that the average chromospheric structure in late-type M stars likely remains constant until field ages, when the TR moves outward toward lower column mass. In future work, we will construct models representative of late-type M stars ($\sim$ 0.1 $M_\sun$) using the \textit{GALEX} measurements from the HAZMAT III sample to guide the models. We will quantify the evolution of EUV flux for M4 -- M8 stars and verify our prediction about the evolution of the temperature structure. 

In addition to the late-type M stars, we will also utilize the \textit{GALEX} measurements from HAZMAT V \citep{Richey2019} to model the evolution of EUV flux emitted from higher-mass K stars and \textit{HST} FUV spectra from HAZMAT IV \citep{Loyd18a} to model M stars in flare states. Future work also includes extending the model temperature structures to include coronal layers, allowing for the computation of the full XUV spectrum. The 1D models presented in this work represent an average for the stellar disk; however, M stars are innately inhomogeneous with spatially varying atmospheric features that correspond with locally enhanced and suppressed regions of short-wavelength emission. Assuming these features are uniformly distributed over the entire disk, an appropriate strategy for modeling the 2D surface would be to combine a synthesis of models with differing weights based on the estimated area of coverage at a specific time in the stellar activity cycle, similar to the solar models in \cite{Fontenla2015}.

\section{Conclusions}\label{sec:conclusion}

Our results quantify the evolution of EUV radiation emitted by early M-type stars (M1 -- M2). We have produced a suite of synthetic EUV -- IR spectra for 0.4 $\pm$ 0.05 $M_\sun$ stars at five ages between 10 Myr and 5 Gyr. This work demonstrates that M stars experience a strong decrease in EUV emission over the main-sequence lifespan. The EUV flux emitted by young M stars is $\sim$100 times greater than that emitted by old M stars and decreases as $\sim t^{-1}$. Between 10 Myr and 5 Gyr, the median fractional EUV flux decreases by a factor of 35. This is consistent with the observed trend of fractional fluxes reducing monotonically with wavelength.

We show that, over time, average M dwarf chromospheres move outward, toward smaller column mass, but that the temperature gradient in the TR remains constant. Similarities between the evolving temperature structures of 0.45 $M_\sun$ and 0.35 $M_\sun$ models and the computed UV spectra suggest that differences of 150 K in $T_{\rm eff}$, 0.1 cm s$^{-2}$ in $g$, and 0.1 $M_\sun$ in $M_{\star}$ do not significantly influence the average structure in the upper atmosphere nor the emergent UV flux. Larger changes in these stellar parameters, however, do impact individual line profiles in the UV spectrum (e.g. an increase of 0.4 cm g$^{-2}$ in log($g$) increases the Ly$\alpha$ line flux by a factor of 1.7).

Differences in the evolutionary trends of the computed Ly$\alpha$ and \ion{Mg}{2} \textit{h} and \textit{k} profiles suggest that \ion{Mg}{2} observations cannot accurately predict the shape of the Ly$\alpha$ line core. Quantitative analysis reveals that the total integrated EUV flux correlates strongly with both FUV and NUV flux as well as Ly$\alpha$ line flux when including the contribution from the continuum. Relative to the continuum, our model Ly$\alpha$ profiles indicate consistency over time. Our models show that the total integrated EUV and FUV flux decrease together at nearly the same rate and that short-wavelength EUV flux correlates strongly with emission lines with high formation temperatures. This study supports the findings of \cite{France2018} that \ion{N}{5} is a good proxy for the 90 -- 360 \AA \ wavelength region.

We present high-resolution synthetic spectra that fill observational gaps for M dwarfs across UV wavelengths, most notably the unobservable EUV spectrum. These spectra provide the wavelength resolution required for input to photochemical and atmospheric escape models, and allow for the study of atmospheric variation over stellar lifetimes. We present a range of spectra with integrated EUV--NUV flux densities that span two orders of magnitude at different ages. At the earliest ages, terrestrial planets located in the canonical M dwarf habitable zone are subject to up to 10,000 times more XUV radiation than the Earth receives at present-day. This increased exposure could negatively impact the potential habitability of these planets.

\acknowledgements

This work was supported by NASA Headquarters under the NASA Earth and Space Science Fellowship Program-grant NNX15AQ94H. T.B. and E.S. acknowledge support from the NASA HST grant HST-GO-14784.001-A and NASA Habitable Worlds grant NNX16AB62G. An allocation of computer time from the UA Research Computing High Performance Computing (HPC) at the University of Arizona is gratefully acknowledged. This work has made use of data from the European Space Agency (ESA) mission {\it Gaia} (\url{https://www.cosmos.esa.int/gaia}), processed by the {\it Gaia} Data Processing and Analysis Consortium (DPAC, \url{https://www.cosmos.esa.int/web/gaia/dpac/consortium}). Funding for the DPAC has been provided by national institutions, in particular the institutions participating in the {\it Gaia} Multilateral Agreement.

\software{PHOENIX \citep{Hauschildt1993,Hauschildt2006,Baron2007}}

\newpage

\appendix
\section{Summary of HAZMAT I Sample}\label{app:minmax}
\begin{table}[ht]
    \centering
    \begin{tabular}{lc|ccc|ccc}
    \hline \hline
      Spectral&Sample&&$F_{\rm FUV}$/$F_{\rm J} \times$ 10$^3$&&&$F_{\rm NUV}$/$F_{\rm J}\times$ 10$^3$&\\
        Type  &Size &Minimum & Median & Maximum & Minimum & Median & Maximum \\
      \hline
      \textbf{10 -- 24 Myr}\\
      \hline
        M1--M4\dotfill &21 \textit{(19) }  & 1.1 & 2.9 & 15.4 & 2.7 & 6.7 & 51.1\\
        M1\dotfill    &6 \textit{(6)}     & 1.1 & 3.7 & 10.9 & 2.7 & 6.3 & 51.1 \\
        M2\dotfill    &7 \textit{(6)}      & 1.1 & 1.8 & 4.8 & 3.3 & 5.5 & 13.6\\
        M3\dotfill    &3 \textit{(3)}         & 6.9 & 15.4 & 15.4 & 14.2 & 21.3 & 21.3\\
        \hline
        \textbf{45 Myr}\\
        \hline
        K7--M4\dotfill &84 \textit{(62)}         & 0.4 & 2.6 & 30.2 & 0.8 & 5.8 & 22.8\\
        M1 &14\dotfill \textit{(14)}             & 0.4 & 2.5 & 5.4 & 0.8 & 6.5 & 10.0\\
        M2&12\dotfill \textit{(11)}                    & 0.5 & 2.4 & 7.4 & 2.6 & 5.6 & 15.4\\
        M3&32\dotfill \textit{(22)}                   & 0.5 & 2.7 & 30.2 & 1.8 & 5.7 & 22.8\\
        \hline
        \textbf{120 -- 300 Myr}\\
        \hline
        K7--M4\dotfill &19 \textit{(18)}              &  0.1 & 2.0 & 4.0 & 0.6 & 4.7 & 11.6\\
        M1\dotfill& 0 &$\cdots$&$\cdots$&$\cdots$&$\cdots$&$\cdots$&$\cdots$\\
        M2\dotfill&8 \textit{(8)}                      & 0.1 & 2.2 & 3.9 & 0.6 & 4.9 & 6.7\\
        M3\dotfill&5 \textit{(4)}                      & 1.5 & 2.9 & 4.0 & 2.9 & 4.0 & 5.4\\
        \hline
        \textbf{650 Myr}\\
        \hline
        M0--M4\dotfill & 30 \textit{(14)}              & 0.1 & 1.1 & 14.0 & 0.3 & 2.6 & 9.3\\
        M1\dotfill & 8 \textit{(3)}             & 0.3 & 0.8 & 4.0 & 0.7 & 1.9 & 5.9\\
        M2\dotfill & 8 \textit{(6)}                      & 0.2 & 1.0 & 3.2 & 0.5 & 3.9 & 6.6\\
        M3\dotfill & 7 \textit{(3)}                     & 0.4 & 3.4 & 14.0 & 0.8 & 5.8 & 8.5 \\
        \hline
        \textbf{$\sim$ 5 Gyr}\\
        \hline
        M0--M3\dotfill & 60 \textit{(34)}               & 0.002 & 0.1 & 2.1 & 0.008 & 0.4 & 4.6 \\
        M1\dotfill &15 \textit{(10)}     & 0.002 & 0.1 & 1.8 & 0.008 & 0.5 & 4.6 \\
        M2\dotfill&17 \textit{(11)}                    & 0.007 & 0.1 & 0.3 & 0.044 & 0.4 & 0.9 \\
        M3\dotfill &23 \textit{(11)}                  & 0.002 & 0.1 & 2.0 & 0.009 & 0.3 & 3.1\\
    \hline
    \end{tabular}
    \caption{Minimum, median, and maximum fractional flux densities (multiplied by 10$^3$ for clarity) from the HAZMAT I \textit{GALEX} sample broken down by age and spectral type. Number of \textit{GALEX} detections (excluding FUV upper limits) is given in italics beside the complete sample size. FUV upper limits are calculated as $(F_{\rm NUV} / F_{\rm J})^{1.11}$.}
    \label{tab:app_ageparams}
\end{table}

\newpage

\section{EUV/Ly$\alpha$ Correlations}\label{app:euvlya}

Estimating EUV emission from exoplanet host stars is critical for studying the atmospheric chemistry of potentially habitable worlds, but it is a difficult problem to solve. Interstellar hydrogen makes observing the 400 -- 912 \AA \ region of the EUV spectrum impossible, and there are very few archival \textit{EUVE} spectra (90 -- 400 \AA) of low-mass stars available that can be used to validate synthetic EUV spectra. Further, the \textit{EUVE} spectra are contaminated by interstellar neutral hydrogen, neutral helium, and ionized helium, and the majority of the EUV emission measured from these stars is below the minimum detectable flux level for the instrument, so only strong emission lines including \ion{He}{2} and highly ionized iron (Fe {\small\rmfamily IX -- XVI \relax}) were identified \citep{craig1997}. However, from these observations, \cite{Linsky2014} showed that the ratio of EUV flux to Ly$\alpha$ flux varies slowly with the Ly$\alpha$ flux, and published formulae for predicting EUV flux ratios in 100 \AA \ wavelength bands. They calculated ratios for wavelengths between 400 and 912 \AA \ using semiempirical models of the Sun from \cite{Fontenla2013}, and ratios for the 912 -- 1170 \AA \ flux from \textit{FUSE} observations of F5 -- M5 stars. 

\begin{figure*}[ht]
    \centering
    \includegraphics[width=1.0\linewidth]{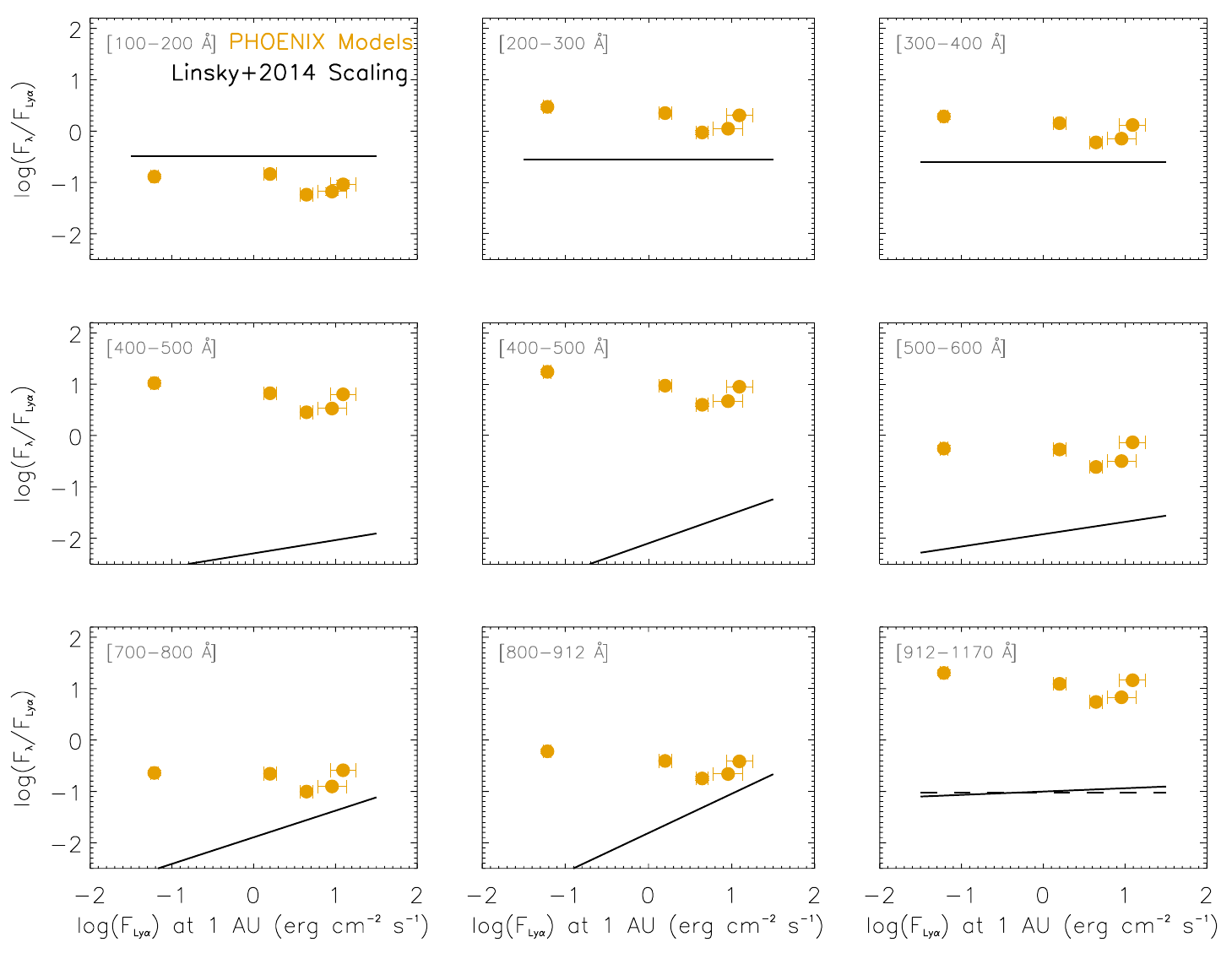}
    \caption{Estimated $F_{\rm \lambda}$/$F_{\rm Ly\alpha}$ using the \cite{Linsky2014} scaling relationships (black) compared to the ratios of integrated fluxes in 100 \AA \ wavelength bands to the Ly$\alpha$ flux computed from the median 0.4 $\pm$ 0.05 $M_\sun$ models (orange). The scaling relationships for wavelengths $<$400 \AA \ are determined from \textit{EUVE} observations of M stars. The relationship for 912 -- 1170 \AA \ is determined from \textit{FUSE} observations of K5 -- M5 stars. The relationships for wavelengths between 400 and 912 \AA \ are computed from solar models from \cite{Fontenla2013}. }
    \label{fig:lyacor}
\end{figure*}

The estimated accuracy of the \cite{Linsky2014} relationships for M stars is 20\% \citep{Youngblood16}. We test the agreement of our synthetic spectra to these scalings in Figure \ref{fig:lyacor}. The integrated EUV fluxes in 100 \AA \ wavelength bands for our median models are listed in Table \ref{tab:m1_10modelsEUV}, along with the continuum-subtracted Ly$\alpha$ line flux. We plot the ratio of $F_{\rm EUV}$/$F_{Ly\alpha}$ versus $F_{Ly\alpha}$ (scaled to 1 AU) in Figure \ref{fig:lyacor} and compare the models to the empirical scaling relations in black. We find that the models follow a similar trend as the scalings derived from \textit{EUVE} observations (100 -- 400 \AA), but deviate significantly at longer wavelengths.

The scaling relationships from 400 to 912 \AA \ are determined from \cite{Fontenla2013} solar models, which may not be appropriate for M stars. \cite{Drake2019} generated synthetic EUV spectra of M dwarfs using differential emission measures (T$_{\rm Tmin}$ = (1--4) $\times$ 10$^5$ K, T$_{max}$ = 10$^6$ -- 10$^7$ K) constrained by \textit{Chandra} X-ray observations. While the synthetic spectra from \cite{Drake2019} carry large uncertainties at wavelengths $>$ 400 \AA, the best-fit model to the \textit{Chandra} data predicts over an order of magnitude higher EUV flux at these wavelengths, consistent with our models. This suggests that the solar EUV spectrum is likely not analogous to M star EUV spectra and highlights differences between our PHOENIX models and the Solar Spectral Radiation Physical Modeling (SRPM) tools \citep{Fontenla2015}.

Our models and the scaling derived from \textit{FUSE} observations (912 -- 1170 \AA) follow the same linear trend, but the modeled $F_{\rm EUV}$/$F_{\rm Ly\alpha}$ fluxes are $\sim$2 orders of magnitude higher. The elevated model fluxes are due to two very strong \ion{O}{6} lines at 1031 and 1038 \AA. The lines are computed in LTE and over estimate the flux. In future work, we will compute models with all species in non-LTE.
 
\begin{table*}[ht]
    \centering
    \begin{tabular}{lccccc}
    \hline \hline
    $\lambda$ Interval (\AA) & 10 Myr & 45 Myr & 120 Myr & 650 Myr & 5000 Myr\\
    \hline
    100 -- 200\dotfill & 0.8	$\pm$	0.4	&	1.1	$\pm$	0.6	&	0.73	$\pm$	0.04	&	0.76	$\pm$	0.05	&	0.03	$\pm$	0.004	\\
    200 -- 300\dotfill  &17.7	$\pm$	6.9	&	16.7	$\pm$	6.7	&	11.9	$\pm$	1.0	&	11.6	$\pm$	1.3	&	0.60	$\pm$	0.01	\\
    300 -- 400\dotfill  &11.4	$\pm$	4.3	&	10.6	$\pm$	4.1	&	7.6	$\pm$	0.7	&	7.4	$\pm$	0.8	&	0.35	$\pm$	0.05	\\
    400 -- 500\dotfill  &54.8	$\pm$	19.5	&	49.4	$\pm$	17.8	&	35.9	$\pm$	3.1	&	34.3	$\pm$	3.9	&	2.10	$\pm$	0.01	\\
    500 -- 600\dotfill  &76.2	$\pm$	25.6	&	68.1	$\pm$	22.9	&	50.3	$\pm$	4.3	&	48.3	$\pm$	5.4	&	3.49	$\pm$	0.02	\\
    600 -- 700\dotfill  &6.4	$\pm$	2.3	&	4.7	$\pm$	1.6	&	3.1	$\pm$	0.4	&	2.8	$\pm$	0.5	&	0.11	$\pm$	0.01	\\
    700 -- 800\dotfill  &2.3	$\pm$	0.9	&	1.9	$\pm$	0.7	&	1.3	$\pm$	0.1	&	1.1	$\pm$	0.2	&	0.05	$\pm$	0.01	\\
    800 -- 900\dotfill &2.6	$\pm$	1.0	&	2.6	$\pm$	1.0	&	1.8	$\pm$	0.1	&	1.6	$\pm$	0.1	&	0.10	$\pm$	0.01	\\
    900 -- 1000\dotfill &4.7	$\pm$	1.6	&	4.4	$\pm$	1.6	&	3.2	$\pm$	0.3	&	3.0	$\pm$	0.3	&	0.21	$\pm$	0.01	\\
    100 -- 1000\dotfill &179.2	$\pm$	63.5	&	162.0	$\pm$	58.3	&	117.7	$\pm$	10.2	&	112.6	$\pm$	12.5	&	7.17	$\pm$	0.03	\\
    Ly$\alpha$\dotfill  & 8.6	$\pm$	3.1	&	14.9	$\pm$	5.9	&	12.7	$\pm$	0.1	&  5.2	$\pm$	0.2	&	0.20	$\pm$	0.03	\\

    \hline
    \end{tabular}
    \caption{EUV fluxes in units of 10$^{5}$ erg s$^{-1}$ cm$^{-2}$ integrated over 100 \AA \ bandpasses for the median models per age, the full 10 -- 100 nm and Lyman alpha flux. All fluxes are taken at the stellar surface. Ly$\alpha$ flux is integrated over 1211.5 -- 1219.7 \AA.}
    \label{tab:m1_10modelsEUV}
\end{table*}

\end{document}